\documentclass[%
 reprint,longbibliography,
showpacs,preprintnumbers,
bibnotes,
 amsmath,amssymb,
prb,
]{revtex4-2}

\usepackage{graphicx}
\usepackage[english]{babel}
\usepackage{microtype}
\usepackage[utf8]{inputenc}
\usepackage[breaklinks=true,colorlinks=true,linkcolor=blue,urlcolor=blue,citecolor=blue]{hyperref}

\usepackage[dvipsnames]{xcolor}      
\usepackage{mathtools}
\mathtoolsset{showonlyrefs=true}
\usepackage{braket}
\usepackage{tikz}

 \usepackage{longtable}
  \usepackage{booktabs}
 \usepackage{siunitx}
 \usepackage{csquotes}

 \usepackage{multirow}
\usepackage{amssymb}

\newcommand{\eqrefs}[2]{Eqs.~\eqref{#1} and \eqref{#2}}
\newcommand{\eqrefss}[3]{Eqs.~\eqref{#1}, \eqref{#2} and \eqref{#3}}

\newcommand{\eref}[1]{Eq.~\eqref{#1}}

\newcommand{\cbar}[1]{\ensuremath{\xbar{#1}}}

\newcommand{\tempbar}[3][\textstyle]{\settowidth{\dimen1}{$#1\bar{#2}$}\makebox[\dimen1][l]{$#1\bar{#2\mspace{#3}}$}}

\newcommand{\xbar}[1]{{\mathchoice{\tempbar[\displaystyle]{#1}{3.5mu}}{\tempbar{#1}{3.5mu}}{\tempbar[\scriptstyle]{#1}{3.5mu}}{\tempbar[\scriptscriptstyle]{#1}{3mu}}}}
\let\inodot\i
\renewcommand{\i}{\mathrm{i}}
\renewcommand{\d}{\mathrm{d}}
\newcommand{\tn}[1]{\textnormal{#1}}

\usepackage{dcolumn}
\usepackage{bm}

\makeatletter
\g@addto@macro\bfseries{\boldmath}
\newcommand*{\balancecolsandclearpage}{%
   \close@column@grid
   \clearpage
   \twocolumngrid
 }
\makeatother

\begin{document}
\preprint{APS/123-QED}

\title{Accelerating Nonequilibrium Green functions simulations with embedding selfenergies}
\author{Karsten Balzer}
\affiliation{Rechenzentrum, Christian-Albrechts-Universität zu Kiel, D-24098 Kiel, Germany}
\author{Niclas Schlünzen}
\affiliation{Center for Advanced Systems Understanding (CASUS), Görlitz, D-02826, Germany }
\author{Hannes Ohldag, Jan-Philip Joost,  and Michael Bonitz
 \email{bonitz@theo-physik.uni-kiel.de}
 }
\affiliation{
Institut f\"ur Theoretische Physik und Astrophysik,
Christian-Albrechts-Universit\"{a}t zu Kiel, D-24098 Kiel, Germany \\ and
Kiel Nano, Surface and Interface Science KiNSIS, Kiel University, Germany
}

\date{\today}

\begin{abstract}
Real-time nonequilibrium Green functions (NEGF) have been very successful to simulate the dynamics of correlated many-particle systems far from equilibrium. However, NEGF simulations are computationally expensive since the effort scales cubically with the simulation duration. Recently we have introduced
the G1--G2 scheme that allows for a dramatic reduction to time-linear scaling  [Schl\"unzen \textit{et al.}, Phys.~Rev.~Lett.~\textbf{124}, 076601 (2020); Joost \textit{et al.}, Phys.~Rev.~B \textbf{101}, 245101 (2020)]. Here we tackle another problem: the rapid growth of the computational effort with the system size. In many situations where the system of interest is coupled to a bath, to electric contacts or similar macroscopic systems for which a microscopic resolution of the electronic properties is not necessary, efficient simplifications are possible. This is achieved by the introduction of an embedding \mbox{selfenergy -- a} concept that has been successful in standard NEGF simulations. Here, we demonstrate how the embedding concept can be introduced into the G1--G2 scheme, allowing us to drastically accelerate NEGF embedding simulations. The approach is compatible with all advanced selfenergies that can be represented by the G1--G2 scheme [as described in Joost \textit{et al.}, Phys.~Rev.~B \textbf{105}, 165155 (2022)] and retains the memory-less structure of the equations and their time-linear scaling.
As a numerical illustration we investigate the charge transfer between a Hubbard nanocluster
and an additional site which is of relevance for the neutralization of ions in matter.
\end{abstract}

\maketitle




The nonequilibrium properties of correlated many-particle systems following a rapid excitation have recently attracted high interest. This applies to many fields such as atoms in optical lattices~\cite{xia_quantum_2015,schluenzen_prb16}, correlated electrons in solids~\cite{jensen_ultrafast_2013}, femtosecond laser pulse excited atoms and molecules~\cite{perfetto_pra_15, lackner_pra_17}, or dense plasmas~\cite{graziani_cpp_21}.
Understanding the nonequilibrium behavior is the basis for potential applications such as ultrafast light-driven electronics~\cite{hommelhoff_nature_2022} or novel material diagnostics using highly charged ions~\cite{niggas_prl_22}.

A theoretical description of correlated fermions far away from equilibrium is very challenging. Among the tools available are wave-function based methods, e.g.~\cite{hochstuhl_epjst_14}, time-dependent density functional theory (TD-DFT), density matrix normalization group (DMRG) simulations, reduced density matrix theory and nonequilibrium Green functions (NEGF) theory. Here, we focus on NEGF simulations because they have undergone a dramatic development during recent years, for an overview, see e.g. the monographs~\cite{stefanucci_nonequilibrium_2013,balzer-book} and the recent reviews by Schl\"unzen \textit{et al.} and Ridley \textit{et al.}~\cite{schluenzen_jpcm_19, ridley_jpa_22}. At the same time, NEGFs are not plagued by many restrictions of other methods and do not exhibit the exponential scaling with the system size known from wave-function-based approaches.

However, NEGF simulations exhibit a very unfavorable cubic scaling of the computation time with the
number of time steps $N_\tn{t}$, which restricts the simulations to very short times. With the restriction to the time diagonal, which is achieved with the generalized Kadanoff-Baym ansatz (GKBA)~\cite{lipavski_prb_86}, the scaling improves to $\mathcal{O}(N_\tn{t}^2)$, but only for the low-order selfenergies, such as the second-order Born approximation. For more accurate approximations ($T$-matrix, $GW$ etc.) the cubic scaling remains. The situation radically changed with the introduction of the G1--G2 scheme by Schl\"unzen \textit{et al.}~\cite{schluenzen_prl_20} which exactly reformulates the GKBA into coupled time-local equations for the one-particle and two-particle Green functions, $G^<(t)$ and  $\mathcal{G}(t)$. This scheme eliminates all memory integrals and, therefore, scales linearly with $N_\tn{t}$. Interestingly, this favorable scaling is achieved already after a small number of time steps and for all common selfenergies, including the second-order Born approximation, the $T$-matrix approximation and $GW$, as was demonstrated by Joost \textit{et al.}~\cite{joost_prb_20}.
The G1--G2 scheme was recently applied to the photoionization of organic molecules~\cite{pavlyukh_prb_21} and ultrafast electron-boson dynamics~\cite{karlsson_prl21}. In particular, G1--G2 simulations with the $GW$ selfenergy were reported for the simulation of ultrafast carrier and exciton dynamics in 2D materials by Perfetto \textit{et al.}~\cite{perfetto_prl_22}.
Despite the importance of $GW$ simulations, they apply only to weakly and moderately coupled many-particle systems. At the same time, recently, many moderately or strongly correlated materials came into the focus of research, including transition-metal dichalcogenides (TMDC) and twisted bilayers of graphene or TMDCs, e.g.~\cite{wu_prl_18, li_nat_21, smolenski_nat_21, bonitz_pj_21}.
Such systems can be treated more accurately via a selfconsistent combination of $GW$ and $T$-matrix diagrams which leads to the dynamically screened ladder approximation (DSL) that has recently been realized within the G1--G2 scheme by Joost \textit{et al.}~\cite{joost_prb_22}.

In G1--G2 simulations, now the memory consumption is the main bottleneck due to the need to store the two-particle Green function $\mathcal{G}$.
Recently, a novel quantum fluctuations approach was presented that eliminates the need to store $\mathcal{G}$ and allows to drastically reduce the memory requirements of nonequilibrium $GW$ simulations~\cite{schroedter_cmp_22}. An alternative approach to reduce the basis dimension is to restrict the simulations to a small number of ``active'' orbitals or degrees of freedom, as done also in time-dependent restricted active-space methods in atomic and molecular physics, e.g.~\cite{hochstuhl_pra_12, hochstuhl_epjst_14} and references therein.

An alternative idea to reduce the dimensionality of the problem is an ``embedding'' approach: here, the \mbox{(sub-)system} of interest is treated with full microscopic detail, whereas its environment, the dynamics of which are of minor importance, is computed in a suitably simplified fashion. Such schemes have been developed in many fields, including quantum chemistry, e.g. the
hybrid quantum mechanics/molecular mechanics approach~\cite{WARSHEL1976227}, condensed matter, e.g. within dynamical mean-field theory~\cite{wael-chibani-phd_16}, the statistical theory of open systems~\cite{ness_pre_14}, and plasma-surface interaction~\cite{Bonitz_fcse_19,bronold_jap_22}. In NEGF simulations, the embedding concept has been successfully applied as well and allows, in particular, for an efficient treatment of nonequilibrium problems and ultrafast electron dynamics. This includes quantum transport in nanoscale junctions coupled to macroscopic leads~\cite{khosvari_prb_12,rabani_jcp_13}, the excitation dynamic of excitonic insulators~\cite{tuovinen_prb_20}, the photoionization of atoms in strong laser fields~\cite{perfetto_pra_15}, or the Auger decay in molecules~\cite{covito_pra_18}, for a text book overview see Ref.~\cite{stefanucci_nonequilibrium_2013}.

In this paper, we extend the NEGF embedding concept to the G1--G2 scheme.
We derive explicit general expressions for the embedding selfenergy and the embedding collision integral, allowing for interaction effects in the environment and the system-environment coupling on the mean-field (Hartree-Fock) level. Compared to the standard G1--G2 scheme, the embedding selfenergy gives rise to an additional equation for the system-environment coupling Green function, $G^{\tn{es},<}$, which is time-local as well. Thus, the resulting equations of motion retain the time-local structure of the equations, for any correlation selfenergy, and thus the favorable time-linear scaling.
As a numerical illustration, we consider the time-dependent charge transfer between a finite Hubbard nanocluster and an additional site which mimicks the neutralization of highly charged ions in matter~\cite{balzer_cpp_21,niggas_prl_22}. We verify good agreement with previous NEGF simulations, as long as the charge transfer is weak. In contrast, for strong charge transfer, deviations arise. We demonstrate how the \mbox{G1--G2} embedding scheme has to be modified in order to restore complete agreement with full two-time NEGF (embedding) simulations.

\section{Nonequilibrium Green functions theory}\label{s:negf}

\subsection{Keldysh-Kadanoff-Baym equations (KBE)}\label{ss:kbe}

Nonequlibrium Green functions theory is formulated in second quantization (for textbook or review discussions, see Refs.~\cite{stefanucci_nonequilibrium_2013,kadanoff-baym-book, schluenzen_cpp16}). For an arbitrary single-particle basis with orbital $\ket{i}$ and spin projection $\sigma$,
one defines creation and annihilation operators, $\hat{c}^{\dagger}_{i\sigma}$ and $\hat{c}_{i\sigma}$, that obey the known anti-commutation rules. These operators are time-dependent via the Heisenberg representation of quantum mechanics. The central quantity of the theory is the one-particle NEGF (we use $\hbar = 1$),
\begin{align}
G_{ij\sigma}(t,t') =
-\i
\langle T_{\cal  C}\hat{c}_{i\sigma}(t)\hat{c}^{\dagger}_{j\sigma}(t')\rangle\,,
    \label{eq:negf}
\end{align}
where the expectation value is computed with the equilibrium density operator of the system. Furthermore, times are running along the Keldysh contour $\cal C$, and $T_{\cal C}$ denotes ordering of operators on $\cal C$, for details see Ref.~\cite{balzer-book}. Referring to observables, for example, the time-dependent electron density in orbital $i$ follows from $G$ via
$\langle\hat{n}_{i\sigma}\rangle(t)= -\i G_{ii\sigma}(t, t^+)$, where $t^+ \equiv t+\epsilon$, with $\epsilon>0$ and $\epsilon \to 0$. If the orbital indices differ, $i\ne j$, the Green function describes time-dependent transitions of electrons between two orbitals. In similar manner one computes the density matrix, currents, mean energies, optical absorption or electrical conductivity from~$G$.

The NEGF obeys the two-time Keldysh-Kadanoff-Baym equations (KBE)~\cite{kadanoff-baym-book},
\begin{align}
 & \sum_k\left[\i\partial_t\delta_{ik}
 -h^{\rm HF}_{ik\sigma}(t)\right]G_{kj\sigma}(t,t')
 \label{eq.kbe}\\
 &\quad\nonumber
 =\delta_{\cal C}(t,t')\delta_{ij}+\sum_{k}\int_{\cal C} \d \bar{t}\,\Sigma_{ik\sigma}(t,\bar{t})G_{kj\sigma}(\bar{t},t')\,,
\end{align}
where $h^{\rm HF}$ contains the one-particle kinetic, potential and mean-field (Hartree-Fock) energy contributions, and correlation effects, on the other hand, are included in the selfenergy $\Sigma$ [we do not consider spin changes and omit the second equation which, is the adjoint of Eq.~\eqref{eq.kbe}].

Without the right hand side, Eq.~\eqref{eq.kbe} would be equivalent to a Vlasov equation or its quantum generalization (time-dependent Hartree-Fock, TDHF). The r.h.s.~contains correlation effects that are responsible for relaxation and dissipation and include scattering of electrons with electrons, ions or lattice vibrations (phonons). Notice the time integral on the r.h.s.~which incorporates memory effects that are important to correctly treat correlations. The standard Boltzmann equation is recovered by evaluating this time integral approximately via a retardation expansion (Markov limit)~\cite{bonitz_qkt, bonitz_cpp18}.

The NEGF formalism is formally exact if the selfenergy would be known exactly. The approach is internally consistent, obeys conservation laws and is applicable to arbitrary length and time scales. Its accuracy is determined by the proper choice for a single function -- the selfenergy. For an overview on the treatment of weak and strong correlations in solids and optical lattices, see Refs.~\cite{schluenzen_cpp16, schluenzen_jpcm_19}.

\subsection{Embedding selfenergy approach}\label{ss:negf-embedding}

In this section, we briefly summarize the nonequilibrium embedding selfenergy approach of NEGF theory following the presentation of Ref.~\cite{Bonitz_fcse_19}, generalizing the methods described in Sec.~\ref{ss:kbe} to open systems. We start from the second-quantized many-body Hamiltonian for the electrons in the entire many-body system and separate it into a ``central'' system system~(s) and its ``environment''~(e) [we denote $\Omega=\{\tn{e},\tn{s}\}$ and do not write the spin index explicitly],
\begin{align}
\label{eq.ham}
\hat{H}_{\textup{total}}(t) = &\sum_{\alpha\beta\in\Omega}\sum_{ij}h^{\alpha\beta}_{ij}(t)\hat{c}^{\alpha\dagger}_i\hat{c}^\beta_j
\nonumber\\
&+\frac{1}{2}\sum_{\alpha\beta\gamma\delta\in\Omega}\sum_{ijkl}w^{\alpha\beta\gamma\delta}_{ijkl}\hat{c}^{\alpha\dagger}_i\hat{c}^{\beta\dagger}_j\hat{c}^{\gamma}_k\hat{c}^{\delta}_l\,.
\end{align}
Here, the operator $\hat{c}^{\alpha\dagger}_i$ ($\hat{c}^{\alpha}_i$) creates (annihilates) an electron in the state $\ket{i}$ of part $\alpha$. The one-particle Hamiltonian, $h(t)=T+V(t)$, contains the electron's kinetic and the (in general, time-dependent) potential energy, whereas $w$ accounts for all possible electron-electron Coulomb interactions within and between the two parts. We underline that ``environment'' is only a notation for a part of the total system that is going to be treated approximately. In most cases of practical interest, the environment is much larger than the system, but it does not necessarily completely enclose the system as a heat bath. Aside from a ``bath'', this part can also describe leads in quantum transport, atomic or molecular energy levels that are not participating in a certain excitation (such as continuum states) or the gas or plasma phase surrounding atoms or a solid.
At the same time this part of the system can be very complex and heterogeneous, consisting of many sub-parts, so the index ``e'' can be a multi-index describing many baths~\cite{Bonitz_fcse_19, balzer_cpp_21}. In this paper, we focus on short-time phenomena. For long-time effects such as thermalization and emergence of irreversibility in the NEGF formalism, see e.g. Refs.~\cite{haug_2008_quantum, galperin_epjst_21}.

We describe the total system~\eqref{eq.ham} by a one-particle nonequilibrium Green function (NEGF) $G^{\alpha\beta}_{ij}(t,t')$, as introduced in Sec.~\ref{ss:kbe}, but here with an additional $2\times2$ matrix structure ($\alpha, \beta=\Omega$),
\begin{align}
\label{eq.negf}
 G^{\alpha\beta}_{ij}(t,t') &=- \i \langle T_C \hat{c}^\alpha_{i}(t)\hat{c}_{j}^{\beta\dagger}(t')\rangle\,,\\
  \rho_{ij}^{\alpha\beta}(t) &= - \i G^{\beta\alpha}_{ji}(t,t^+)\,,
\label{eq:g-dm}
\end{align}
e.g., Refs.~\cite{stefanucci_nonequilibrium_2013, balzer-book}, and the time-diagonal elements provide the density matrix \eqref{eq:g-dm}, as discussed in Sec.~\ref{ss:kbe}.
The diagonal elements $\rho_{ij}^{\tn{ss}}$ ($\rho_{ij}^{\tn{ee}}$) refer to the system part (to the environment part). Moreover, the density matrix component $\rho_{ij}^{\tn{es}}$ is related to charge and energy transfer processes between  system and environment and will be of special interest in the following.

The equations of motion for the NEGF are the generalization of Eq.~\eqref{eq.kbe}
to the total system (we use Einstein's convention and imply summation over repeating orbital indices $k$)
\begin{align}
 &\i\partial_tG^{\alpha\beta}_{ij}(t,t') - \sum_{\delta=\tn{e},\tn{s}} h^{\rm HF, \alpha\delta}_{ik}(t)G^{\delta\beta}_{kj}(t,t')
 \label{eq.kbe_interface}\\
 &\qquad
 =\delta^{\alpha\beta}_{ij}\delta_C(t,t')+\sum_{\delta=\tn{e},\tn{s}} \int_C\!\!\!\d\bar{t}\,\Sigma^{\alpha\delta}_{ik}(t,\bar{t})G^{\delta\beta}_{kj}(\bar{t},t')\,.\nonumber
\end{align}
In many cases, a full quantum-mechanical treatment of the entire system, including the many degrees of freedom of the environment, is neither possible nor necessary. In the following, we show how it is possible to derive approximate equations for the dynamics of the electrons in the system that still incorporate the leading order influences of the environment. While this ``embedding'' approach is based on a formal decoupling of the system and the environment parts of the KBE, it retains one-electron charge and energy transfer in the single-particle Hamiltonian $h^{\rm HF, se}$, cf.~Eq.~\eqref{eq.sigma.ct} below.

Let us start with the key approximation that correlations in the environment part are negligible, i.e., we will assume $\Sigma^{\tn{ee}}=0$. We still retain interaction effects on the mean-field level. Furthermore, for a macroscopic environment it is reasonable to expect that the coupling to the system is irrelevant for the dynamics of the environment, which allows us to neglect the Hamiltonian $h^{{\rm HF},\tn{es}}_{ik}$ in the equation for $G^{\tn{ee}}$ and to set $\Sigma^{\tn{se}}=\Sigma^{\tn{es}}=0$.
Then, the KBE~\eqref{eq.kbe_interface} for the system, the environment and the cross parts simplify to (we denote $G^{\tn{ss}} \to G^{\tn{s}}$, $G^{\tn{ee}} \to g^{\tn{e}}$ [we reserve the notation $G^{\tn{e}}$ for a different quantity, see Sec.~\ref{ss:problem-fix} below],
 $h^{\rm HF,\alpha \alpha} \to h^{\rm HF,\alpha}$ and $\Sigma^{\alpha \alpha} \to \Sigma^\alpha$)
\begin{align}
\left\{\i\partial_t\delta_{ik}-h^{{\rm HF},\tn{s}}_{ik}(t)\right\}G^{\tn{s}}_{kj}(t,t')&
\label{eq:gss-equation}\\
\nonumber
\qquad
=h^{{\rm HF},\tn{se}}_{i\,\underline k}(t)G^{\tn{es}}_{\underline k \,j}(t,t')
&  +\delta_{ij}\delta_C(t,t') \\
\nonumber
  + \int_C\!\!\!\d\bar{t}\,\Sigma^{\tn{s}}_{ik}(t,\bar{t})&G^{\tn{s}}_{kj}(\bar{t},t')\,,\\
  \left\{\i\partial_t\delta_{\underline i \,\underline k}-h^{{\rm HF},\tn{e}}_{\underline i \,\underline k}(t)\right\}G^{\tn{es}}_{\underline k \,j}(t,t')&=h^{{\rm HF},\tn{es}}_{\underline i\, k}(t)G^{\tn{s}}_{kj}(t,t')\,,
  \quad\label{eq:ges-equation}\\
  \left\{\i\partial_t\delta_{\underline i \,\underline k}-h^{{\rm HF},\tn{e}}_{\underline i \,\underline k}(t)\right\}g^{\tn{e}}_{\underline k\,\underline j}(t,t')&=\delta_{\underline i\, \underline j}\delta_C(t,t')\,.
  \label{eq:gee-equation}
\end{align}
Here and in the following, we will use underlined indices for the orbitals in the environment, for a better distinction.

The NEGF $g^{\tn{e}}$ fulfills a simple isolated Hartree-Fock dynamics, whereas $G^{\tn{es}}$, in addition, is affected by the Hartree-Fock renormalization of the system-environment coupling. We immediately recognize that the equations for $g^{\tn{e}}$ and $G^{\tn{es}}$ contain the same term on the left hand side (parentheses) which is nothing but the inverse Green function $g^{\tn{e}\,-1}_{\underline i\, \underline k}$. Thus, multiplying \eref{eq:ges-equation} by $g^{\tn{e}}_{\underline l\, \underline i}(t,t')$ and integrating over the time contour, we obtain an explicit solution for $G^{\tn{es}}$:
\begin{align}
    G^{\tn{es}}_{\underline l\,j}(t,t') = \int_C\!\!\!\d\bar{t}\, g^{\tn{e}}_{\underline l\,\underline i}(t,\bar t)\, h^{{\rm HF},\tn{es}}_{\underline i\,k}(\bar t)\,G^{\tn{s}}_{kj}(\bar t,t')\,.\label{eq:ges-solution}
\end{align}
Equation~\eqref{eq:ges-solution} allows us to eliminate $G^{\tn{es}}$ from the equation for $G^{\tn{s}}$ and to rewrite this term in the form of an additional selfenergy, $\Sigma^{\rm emb}$:
\begin{align}
\label{eq.kbe.embedding}
&\left\{\i\partial_t\delta_{ik}-h^{{\rm HF},\tn{s}}_{ik}(t)\right\}G^{\tn{s}}_{kj}(t,t')
 \\\nonumber
&=\delta_{ij}\delta_C(t,t')
 +\int_C\!\!\!\d\bar{t}\,\left\{\Sigma^{\rm emb}_{ik}(t,\bar{t})+\Sigma^{\tn{s}}_{ik}(t,\bar{t})\right\}G^{\tn{s}}_{kj}(\bar{t},t')\,.
\end{align}
This embedding selfenergy is given by
\begin{align}
\label{eq.sigma.ct}
  \Sigma^{\rm emb}_{ij}(t,t') &=
  h^{{\rm HF},\tn{se}}_{i\,\underline k}(t)g_{\underline k \, \underline l}^{\tn{e}}(t,t') h^{{\rm HF},\tn{es}}_{\underline l \,j}(t')\,,
\\
h^{\rm HF, \tn{se}}_{i\,\underline j}(t) &= \int\!\!\d^3r\,\phi^{\tn{s}*}_i(\mathbf{r})[\hat{T}+\hat{V}^{\rm HF}(t)]\chi_{\underline j}^{\tn{e}}(\mathbf{r};t)\,,
\label{eq:hsp}
\end{align}
and involves the system-environment coupling Hamiltonian $h^{\rm HF,\tn{se}}$, which is renormalized by the Hartree-Fock mean field.

The KBE~\eqref{eq.kbe.embedding} shows how the many-body description of an isolated (but correlated) system is altered in an open system, i.e.~by the presence of the environment: the electronic states of the environment give rise to an additional selfenergy, $\Sigma^{\textup{emb}}(t,t')$, that renormalizes the energy spectrum of the system. While, for $\Sigma^{\textup{emb}}=0$, Eq.~\eqref{eq.kbe.embedding} conserves the particle number [assuming a conserving approximation for  $\Sigma^{\tn{s}}$, such as Hartree-Fock, second-order Born or $GW$], the inclusion of the embedding selfenergy, in general, explicitly gives rise to time-dependent changes of the particle number and energy in the system. This issue is discussed in more detail in Sec.~\ref{ss:problem-fix}.

For the practical solution of \eref{eq.kbe.embedding}, the coupling Hamiltonian $h^{{\rm HF},\tn{se}}(t)$ has to be computed by selecting the relevant electronic transitions between system and environment and computing the matrix elements of the kinetic and HF-renormalized potential energy operators, $\hat{T}$ and $\hat{V}^{\rm HF}$, with the electronic single-particle wave functions $\phi^{\tn{s}}$ ($\chi^{\tn{e}}$) in the system (environment).

\subsection{Energy and particle transfer between system and environment}\label{ss:g-transfer}

While the equation of motion for $G^{\tn{s}}$ approximately includes the overall influence of the environment via the embedding selfenergy, there is no information visible how different orbitals of the environment contribute. This information is contained in the two-time structure of the embedding selfenergy, Eq.~\eqref{eq.sigma.ct}. However, it enters the equation for $G^{\tn{s}}$ in such a way that all orbitals of the environment are traced out. Nevertheless, our approach allows for reconstructing orbital-resolved properties by analyzing the Green function $G^{\tn{es}}$, Eq.~\eqref{eq:ges-solution}.
From this equation we get the correlation function
\begin{align}
    G^{\tn{es} <}_{\underline l\,j}(t,t') =&
    \int\!\!\d\bar{t}
    \, \bigg\{
    g^{\tn{e,R}}_{\underline l\, \underline i}(t,\bar t)\, h^{{\rm HF},\tn{es}}_{\underline i\,k}(\bar t)\,G^{\tn{s} <}_{kj}(\bar t,t')
    \nonumber\\
    &
    -g^{\tn{e} <}_{\underline l\, \underline i}(t,\bar t)\, h^{{\rm HF},\tn{es}}_{\underline i\,k}(\bar t)\,G^{\tn{s,A}}_{kj}(\bar t,t')
    \bigg\}
    \,,\label{eq:ges-less}
\end{align}
and can compute orbital resolved expectation values of a single-particle observable that couples system and environment, $\hat A= \sum_{\underline i\, j} A^{\tn{es}}_{\underline i\,j}\hat{c}_{\underline  i}^{\tn{e}\dagger}\hat{c}_j^{\tn{s}}$, by tracing over the system states,
\begin{align}
    \langle \hat{A}_{\underline l}\rangle(t) = 
    \pm\i\sum_j A^{\tn{es}}_{\underline l\, j} \, G^{\tn{es} <}_{\underline l\,j}(t,t)\,.
    \label{eq:transfer-average}
\end{align}

\section{Time-local HF-GKBA equations (G1--G2 scheme)}\label{s:g1-g2}

\subsection{General equations}

To derive the form of the embedding selfenergy in the HF-GKBA of Green functions theory, we start from the two-time equation of the system part, Eq.~\eqref{eq.kbe.embedding}, and take the ``less'' component which involves the retarded and less component of the two selfenergy contributions~\cite{stefanucci_nonequilibrium_2013,balzer-book} (we also skip the superscripts ``s'' in the following),
\begin{align}
&\left\{\i\partial_t\delta_{ik}-h^{{\rm HF}}_{ik}(t)\right\}G^{<}_{kj}(t,t')
 \nonumber\\
& =\int\!\!\d\bar{t}\,\left\{\Sigma^{\tn{R}}_{ik}(t,\bar{t})G^{<}_{kj}(\bar{t},t') - \Sigma^{<}_{ik}(t,\bar{t})G^{\tn{A}}_{kj}(\bar{t},t')\right\} \nonumber\\
& +\int\!\!\d\bar{t}\,\left\{\Sigma^{{\rm emb}, \tn{R}}_{ik}(t,\bar{t})G^{<}_{kj}(\bar{t},t') - \Sigma^{{\rm emb}, <}_{ik}(t,\bar{t})G^{\tn{A}}_{kj}(\bar{t},t')\right\}
\,.
\label{eq.kbe_gless-ss}
\end{align}
Computing the difference of this equation and its adjoint, the equation of motion for the single-particle Green function $G^<(t)=G^{\tn{s},<}(t,t)$ on the time diagonal (first equation of the G1--G2 scheme) becomes
\begin{align}
&    \i \partial_t G^{<}_{ij}(t) -
     \left[
    h^{\rm HF},G^{<}\right]^s_{ij,t} =
    \left(I(t)+ I^\dagger(t)\right)_{ij}\,,
    \label{eq:time-diagonal-gless-equation}
    \\
&    I_{ij}(t) = I^{\rm cor}_{ij}(t)+I^{\rm emb}_{ij}(t)\,,\\
&    I^{\rm cor}_{ij}(t) = \int_{t_0}^t \d\bar t
    \left\{\Sigma^{>}_{ik}(t,\bar{t})G^{<}_{kj}(\bar{t},t) - \Sigma^{<}_{ik}(t,\bar{t})G^{>}_{kj}(\bar{t},t)\right\}\,,
\label{eq:collision-integral}
\\
&    I^{\rm emb}_{ij}(t) = \int_{t_0}^t \d\bar t
    \bigg\{\Sigma^{\rm emb, >}_{ik}(t,\bar{t})G^{<}_{kj}(\bar{t},t)
    \nonumber\\
&\qquad\qquad\qquad \qquad   -\Sigma^{\rm emb, <}_{ik}(t,\bar{t})G^{>}_{kj}(\bar{t},t)\bigg\}\,,
\label{eq:collision-integral-emb}
\end{align}
where we introduced the short notations
\begin{align}
    \left[
     A,B\right]^\alpha_{ij,t} &= (AB)^\alpha_{ij,t} - (BA)^\alpha_{ij,t}\,,\quad \alpha=\tn{s}, \tn{e}\,,
     \label{eq:def-commutator}\\
     (AB)^\tn{s}_{ij,t} &= \sum_{k\in \tn{s}} A_{ik}(t)B_{kj}(t)\,,
     \label{eq:def-product}\\
     (AB)^\tn{e}_{ij,t} &= \sum_{\underline k\in \tn{e}} A_{i\,\underline k}(t)B_{\underline k\, j}(t)\,,
\end{align}
where the superscript indicates the sub-space over which the internal summation is performed.

Note that Eq.~\eqref{eq:time-diagonal-gless-equation} is not closed for $G^<(t)$, but still involves two-time functions under the integral. This problem will be solved via the generalized Kadanoff-Baym ansatz (GKBA) in the next section.

\subsection{HF-GKBA approach to the embedding selfenergy}

Applying the Hartree-Fock GKBA~\cite{lipavski_prb_86, hermanns_prb14} allows us to eliminate the two-time functions in Eq.~\eqref{eq:collision-integral} away from the time-diagonal, according to
\begin{align}
    G^\gtrless_{ij}(t,t') =
\i
\left[G_{ik}^\mathrm{R}(t,t') G_{kj}^\gtrless(t')-G_{ik}^\gtrless(t) G_{kj}^\mathrm{A}(t,t')\right]\,,
\label{eq:GKBA}
\end{align}
where the retarded and advanced Green functions are approximated by Hartree-Fock Green functions of the system [Note that there are different versions of the GKBA possible. This issue will be addressed in Sec.~\ref{ss:problem-fix}, where we discuss an extension of the embedding scheme]. With the above ansatz the correlation part of the selfenergy, $\Sigma^{\tn{s},\gtrless}$ can be eliminated, giving rise to an equation of the time-local correlation part of the two-particle Green function $\mathcal{G}(t)$~\cite{schluenzen_prl_20,joost_prb_20},
\begin{align}
    \i \partial_t \mathcal{G}_{ijkl}(t) &- \Big[ h^{(2),\tn{HF}}(t),\mathcal{G}(t) \Big]_{ijkl} = \Psi^\pm_{ijkl}(t)
 \, ,\quad  \label{eq:G1-G2_soa}\\
  h^{(2),\tn{HF}}_{ijkl}(t) &= h^\tn{HF}_{ik}(t)\delta_{jl} + h^\tn{HF}_{jl}(t)\delta_{ik}\,,
 \label{eq:h2-hf}\\
     \Psi^\pm_{ijkl}(t) &= \i^2\sum_{pqrs} w^\pm_{pqrs}(t)\left\{
    \mathcal{G}^{\tn{H},>}_{ijpq} \mathcal{G}^{\tn{H},<}_{rskl}
    - (>\leftrightarrow <)
    \right\}_t\,,
    \label{eq:psi-pm-def}\\
    \mathcal{G}^{\tn{H},\gtrless}_{ijkl}(t) &\coloneqq G^\gtrless_{ik}(t,t) G^\gtrless_{jl}(t,t)\,,
    \label{eq:g2h-def}
\end{align}
where we introduced the (anti-)symmetrized interaction, $w^\pm_{pqrs}=w_{pqrs}\pm w_{pqsr}$ [the time dependence arises from the preparation of the correlated initial state via ``adiabatic switching'']. Furthermore, Eq.~\eqref{eq:G1-G2_soa} is given for the case of the second-order Born selfenergy. The extension to more advanced selfenergies has been presented in Ref.~\cite{joost_prb_22}.
The equation is however not affected by the embedding selfenergy which appears as an additional contribution to the collision integral, on the r.h.s.~of the equation for $G^<$, Eq.~\eqref{eq:time-diagonal-gless-equation}. Thus, our results for the embedding selfenergy are compatible with any correlation selfenergy.

We now demonstrate that, as the collision integral $I^{\rm cor}$, also  the non-Markovian collision integral $I^{\rm emb}$ of Eq.~\eqref{eq:collision-integral-emb} can be transformed into a
time-local expression. First, using the definition of the embedding selfenergy \eqref{eq.sigma.ct}, we write the embedding collision integral as
\begin{align}
    I^{\rm emb}_{ij}(t) =& \int_{t_0}^t \d\bar t
    \bigg\{
    h^{{\rm HF},\tn{se}}_{i\,\underline l}(t)\,g_{\underline l\, \underline m}^{\tn{e} >}(t,\bar t)\, h^{{\rm HF},\tn{es}}_{\underline m\,k}(\bar t)
    G^{<}_{kj}(\bar{t},t)
    \nonumber\\
&  -  h^{{\rm HF},\tn{se}}_{i \, \underline l}(t)\,g_{\underline l\, \underline m}^{\tn{e} <}(t,\bar t)\, h^{{\rm HF},\tn{es}}_{\underline m\, k}(\bar t) G^{>}_{kj}(\bar{t},t)\bigg\}\,,
\label{eq:iembedding-gkba}
\end{align}
where $g^{\tn{e},\gtrless}$ are Hartree-Fock Green functions of the environment that are explicitly known from Eq.~\eqref{eq:gee-equation}. Second, we separate the Hartree-Fock Hamiltonian that is not under the time integral,
\begin{align}
I^{\tn{emb}}_{ij}(t) &= h^{\tn{HF,se}}_{i\,\underline k}(t) G^{\tn{es},<}_{\underline k\, j}(t)\,, \label{eq:iemb-ges}\\
    G^{\tn{es},<}_{\underline i\, j}(t) &= \int_{t_0}^t \d \cbar{t} \, h^{\tn{HF,es}}_{\underline k\, l}(\cbar{t})\Big[g^{\tn{e},>}_{\underline i\,\underline k}(t,\cbar{t}) G^<_{lj}(\cbar{t},t) \\
    & \qquad \qquad \qquad- g^{\tn{e},<}_{\underline i\,\underline k}(t,\cbar{t}) G^>_{lj}(\cbar{t},t)\Big]\,.
\label{eq:ges-integral}
\end{align}
Differentiating $G^{\tn{es},<}$ with respect to time results in two terms. The first is due to differentiation of the upper integration boundary,
\begin{align}
\i \left[\frac{\d }{\d t}G^{\tn{es},<}_{\underline i\,j}(t)\right]_{\int} &=
\i h^{\tn{HF,es}}_{\underline k\,l}(t)\bigg[g^{\tn{e},>}_{\underline i\,\underline k}(t,t) G^<_{lj}(t,t)
\nonumber\\
& \qquad \qquad- g^{\tn{e},<}_{\underline i\,\underline k}(t,t) G^>_{lj}(t,t)\bigg]\nonumber\\
&= \left(h^{\tn{HF,es}} G^<\right)^\tn{s}_{\underline i\, j,t} - \left(g^{\tn{e},<} h^{\tn{HF,es}}\right)^\tn{e}_{\underline i\, j,t} \,,
\nonumber
\end{align}
whereas the second arises from the time dependence of the integrand in Eq.~\eqref{eq:ges-integral},
\begin{align}
&\i \left[\frac{\d}{\d t}G^{\tn{es},<}_{\underline i\,j}(t)\right]_{t}\\
&= \int_{t_0}^t \d \cbar{t} \, h^{\tn{HF,es}}_{\underline k\,l}(\cbar{t})\Big[h^{\tn{HF,e}}_{\underline i\,\underline m}(t) g^{\tn{e},>}_{\underline m\,\underline k}(t,\cbar{t}) G^<_{lj}(\cbar{t},t) \\[-0.5pc]
&\qquad\qquad\qquad\qquad\quad
- g^{\tn{e},>}_{\underline i\,\underline k}(t,\cbar{t}) G^<_{lm}(\cbar{t},t) h^{\tn{HF}}_{mj}(t) \nonumber \\
&\qquad\qquad\qquad\qquad\quad
- h^{\tn{HF,e}}_{\underline i\,\underline m}(t) g^{\tn{e},<}_{\underline m\,\underline k}(t,\cbar{t}) G^>_{lj}(\cbar{t},t) \\
&\qquad\qquad\qquad\qquad\quad
+ g^{\tn{e},<}_{\underline i\,\underline k}(t,\cbar{t}) G^>_{lm}(\cbar{t},t) h^{\tn{HF}}_{mj}(t)\Big]
\nonumber\\[0.5pc]
&= \left(h^{\tn{HF,e}} G^{\tn{es},<}\right)^\tn{e}_{\underline i\,j,t} - \left(G^{\tn{es},<} h^{\tn{HF}}\right)^\tn{s}_{\underline i\, j,t}\,,
\nonumber
\end{align}
where, in the differentiation of $G^<$, the HF-GKBA was used.
Collecting the two terms together, we finally obtain
\begin{align}
\i \frac{\d}{\d t} G^{\tn{es},<}_{\underline i\,j}(t) &
- \left\{ \left(h^{\tn{HF,e}} G^{\tn{es},<}\right)^\tn{e}_{\underline i\, j,t} -
\left(G^{\tn{es},<} h^{\tn{HF}}\right)^\tn{s}_{\underline i\, j,t}
\right\}\\
&= \left(h^{\tn{HF,es}} G^<\right)^\tn{s}_{\underline i\, j,t} - \left(g^{\tn{e},<} h^{\tn{HF,es}}\right)^\tn{e}_{\underline i\, j,t}\,.
\qquad
\label{eq:ges-equation-local}
\end{align}
Here, the l.h.s.~contains the single-particle  (Hartree-Fock) dynamics of $G^{\tn{es}}$, whereas the r.h.s.~can be understood as inhomogeneity, which is a consequence of the coupling of $G^{\tn{es}}$ to the Green functions of the system and the environment, respectively.

With this we have succeeded in deriving a time-local equation of motion for the Green function that couples our system to the environment. Inserting the solution of this equation into Eq.~\eqref{eq:iemb-ges}, the embedding collision integral can be computed and inserted into the equation of motion for $G^<$, Eq.~\eqref{eq:time-diagonal-gless-equation}, which closes the G1--G2 scheme for the case of an embedding selfenergy.

\section{Numerical example and further improvement of the G1--G2 scheme }
\label{s:example}

\subsection{Time-dependent charge transfer model between a finite Hubbard cluster and its environment}

We consider a finite Hubbard nanocluster which is coupled to external sites or orbitals that represent the ``environment''. This can be considered as a prototype model for current flow between a correlated material and external leads or for resonant charge transfer between a correlated target and an impacting ion. In fact, a NEGF embedding selfenergy approach was recently presented for the latter case for finite graphene clusters in Ref.~\cite{balzer_cpp_21} and extended to monolayers of graphene and MoS$_2$ in Ref.~\cite{niggas_prl_22}. Here, we use the same model and apply it to the present G1--G2 scheme. This allows us to compare the G1--G2 results to known NEGF benchmark data.

To simplify the model, we consider interactions on the Hartree-Fock level, i.e.
\begin{align}
h^{\tn{HF},\tn{s}}_{ij}(t)=-J\delta_{\langle i,j\rangle}+\delta_{ij}\,U\left( \langle \hat{n}^{\tn{s}}_i\rangle(t)-\frac{1}{2}\right)\,,
\label{eq:hf-ham}
\end{align}
where $J$ is the nearest-neighbor hopping constant ($\delta_{\langle i,j\rangle}=1$ for nearest neighbors and zero otherwise), $U$ denotes the on-site Hubbard interaction strength, and $\langle \hat{n}_i^{\tn s}\rangle(t)=-\i G^{\tn{s},<}_{ii}(t)$. Note that we drop any spin indices as the system is assumed to be throughout in the paramagnetic state. Furthermore, the nanocluster couples to one additional environment site (with index ``0'') via the lattice site ``1'', and for the system-environment coupling we apply the model of Ref.~\cite{balzer_cpp_21},
\begin{align}
h^{\tn{se}}_{i0}(t)&=\delta_{i1}\gamma(t)\,,\\
\gamma(t)&=\gamma_0 \exp[-(t-t_\gamma)^2/(2\tau_\gamma^2)]\,,
\end{align}
which was found to reproduce the charge transfer between a highly charged ion impacting graphene monolayers very well.
In case of highly charged ions, this charge transfer can be very intense (depending on the ion charge) and rapid (depending on the ion velocity). Both properties can be directly controlled by the amplitude $\gamma_0$ and the pulse duration $\tau_\gamma$.
 In the numerical simulations, we measure energies in units of $J$, and times in units of $t_0=\hbar J^{-1}$.

To simplify the situation even further, here we concentrate on a finite Hubbard chain of length $L$, which is prepared in the ground state at half-filling  [$\langle \hat{n}_{i}^{\tn{s}}\rangle(0)=0.5$ with $i=1,\ldots, L$; coupling to the environment site at the chain's one end], and
choose $h^{\tn{e}}_{00}(t)=\epsilon$ and $n_0^{\tn e}=\langle \hat{n}_0^{\tn e}\rangle(0)$ to be the energy and initial occupation of the environment site ``0'', respectively. We performed extensive simulations for various system sizes and excitation conditions. The main results can be summarized as follows: for weak charge transfer (small $\gamma_0$) the present G1--G2 embedding scheme exhibits very good agreement with the previous NEGF embedding results. However, for $\gamma_0 \gtrsim 0.5J$, noticeable deviations are found that increase with $\gamma_0$. Charge transfer may even lead to negative site occupations which is, of course, unphysical. No such behavior is observed in the two-time simulations, for identical model parameters.

\subsection{Analyzing and fixing the problems of the G1--G2 embedding scheme}\label{ss:problem-fix}

\begin{figure}[h]
\centering
\includegraphics[width=0.4825\textwidth]{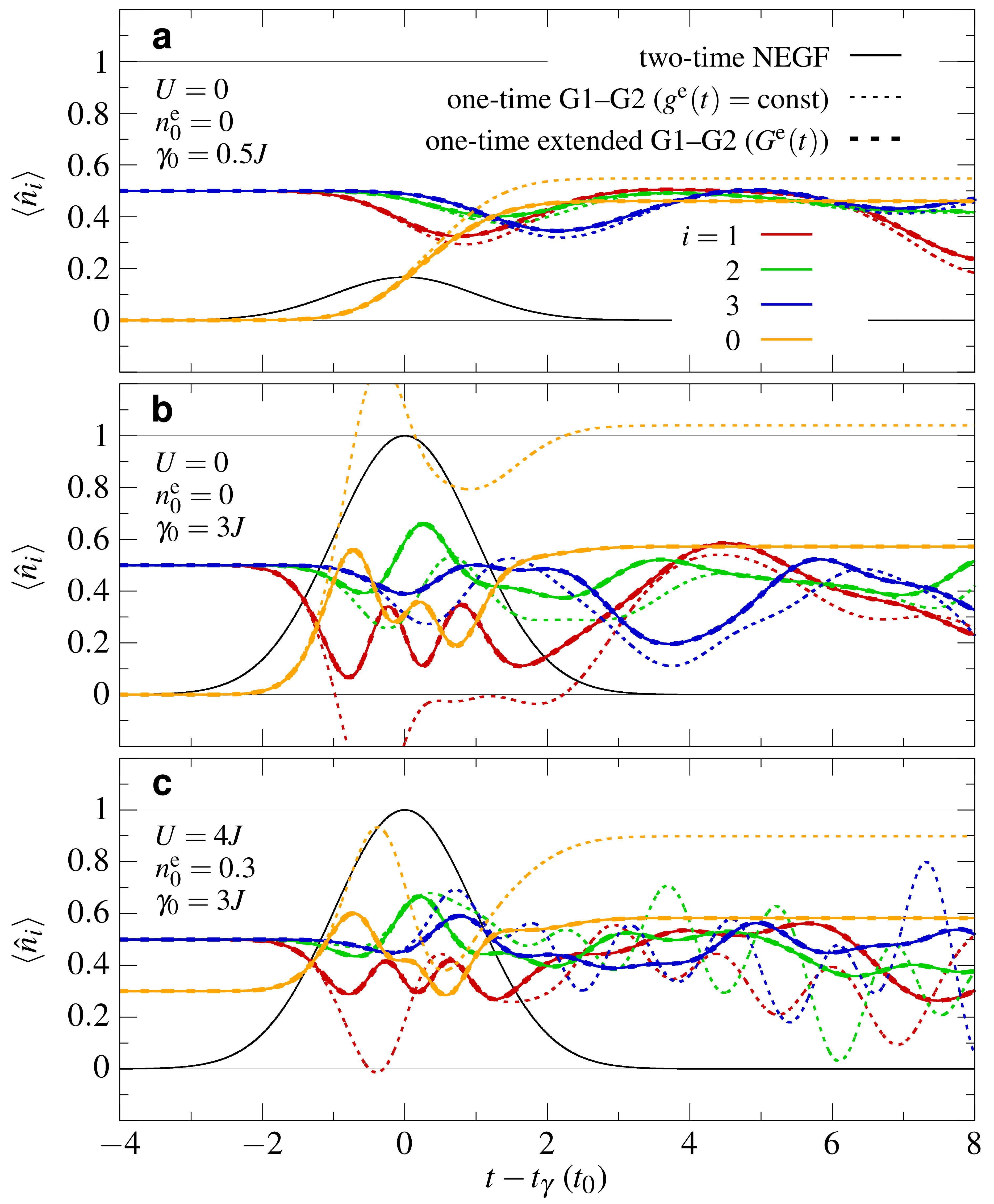}
\caption{Strong charge exchange between a six-site Hubbard chain and a single site with index``0''. Shown are the time-dependent electron densities on the three sites ``1'', ``2'', ``3'' of the chain, that are adjacent to the external site. Initially, the chain is at half-filling. (a) and (b): The lattice electrons are non-interacting ($U=0$), and the additional site is empty, $n^\tn{e}_0=0$. (c): Interaction of the electrons is treated on the Hartree-Fock level with $U=4J$ and $n^\tn{e}_0=0.3$. The black solid lines indicate the function $\gamma(t)$ [scaled by a factor $1/3$] with the pulse width $\tau_\gamma=1t_0$ in all panels. The amplitude equals   $\gamma_0=0.5J$ (a) and $\gamma_0=3J$, (b) and (c). Three sets of results are shown: two-time NEGF embedding results (full lines), the G1--G2 model of Sec.~\ref{s:g1-g2} (dots) and the extended embedding model, Eqs.~\eqref{eq:ges-new} and~\eqref{eq:gee-new} (dashes).
}
\label{fig:6sites}
\end{figure}

Let us analyze the problems of the G1--G2 scheme in more detail. To this end, we study the case of just $L=6$ Hubbard sites. We vary the intensity and duration of the charge transfer in broad ranges so that the model covers realistic situations of highly charged ion experiments~\cite{niggas_prl_22}.
Three examples are shown in Fig.~\ref{fig:6sites}.
Panels (a) and (b) refer to the simplest case of a non-interacting Hubbard chain ($U=0$) coupled to an initially empty site, $n^\tn{e}_0=0$. Consider first part (a), where the amplitude is moderate, $\gamma_0=0.5J$. During the pulse $\gamma(t)$ [cf.~the black lines peaking at time $t=t_\gamma$], the density at site ``1'' (red) decreases, followed by a delayed and weaker depletion of sites ``2'' and ``3''. Simultaneously the occupation of the external site (``0'', yellow) increases, reaching about half-filling. Notice that there are two sets of curves: full lines refer to two-time NEGF embedding simulations whereas the dotted lines refer to the present G1--G2 embedding scheme. In Fig.~\ref{fig:6sites}a, there is overall good agreement between both simulations. The largest deviations are observed in the density $\langle \hat{n}^\tn{e}_0\rangle(t)$ which are of the order of $20\%$.

The situation dramatically changes in Fig.~\ref{fig:6sites}b, where we increase the amplitude to $\gamma_0=3J$. Consider first the two-time NEGF simulations (full lines).
Already before the peak of $\gamma(t)$ the site nearest to the external one (i.e. site ``1'') is almost completely depleted whereas the external site exceeds half filling. After the pulse has passed, the density $\langle \hat{n}^\tn{e}_0\rangle(t)$ remains almost constant whereas the site occupations of the chain continue to exhibit nonlinear oscillations. Note that the two-time embedding simulations are easily tested: to this end we have performed NEGF simulations for the total system, including the additional site, i.e. for a $7$-site chain (avoiding the embedding concept). The agreement is perfect in all cases we considered.

Consider now the G1--G2 results (dotted lines). Initially, for small $\gamma(t)$ the densities are in good agreement with the two-time results. However, when the excitation reaches about half of the maximum value, the two results start to differ qualitatively: the density on site ``1'' becomes negative whereas site ``0'' is more than doubly occupied. Such unphysical behavior persists for the entire duration of the simulation. Similar behavior was observed in many other situations of strong charge transfer (large $\gamma_0$).
For illustration, another example is shown in Fig.~\ref{fig:6sites}c. There, we kept the same $\gamma_0$, but considered an increased initial occupation, $n^\tn{e}_0=0.3$ and also included interaction effects in the chain on the Hartree-Fock level ($U=4J$). While the problem of densities outside the allowed range is reduced, the deviations from the two-time results are striking as well.
We verified that the observed problems are not numerical artifacts, but must be a rooted in the present G1--G2 embedding model.

So what is wrong? The answer is simple: when solving Eq.~\eqref{eq:ges-equation-local}, the present model does not take into account the time evolution of the density on the additional site ``0''; this density is assumed to be constant, cf.~Eq.~\eqref{eq:gee-equation}. This assumption is certainly justified in typical ``embedding'' situations where the central system is coupled to a very large environment with many degrees of freedom which is not modified by the system. In the present case, however, we considered a completely different situation where the environment is represented by a single orbital (site ``0''), the occupation of which changes significantly during the interaction with the system which is very strong. In this case, obviously, Pauli blocking and ``population inversion'' of sites ``0'' and ``1'' should be expected to become relevant.
 Since the latter situation of very strong and spatially localized excitation is a case of direct experimental relevance in the interaction of highly charge ions with matter \cite{niggas_prl_22}, it would be desirable to extend the G1--G2 embedding scheme to the case of very strong coupling. In the following, we present the solution to this task and demonstrate how to eliminate the observed problems.

 To this end, we return to Eqs.~\eqref{eq:gss-equation}--\eqref{eq:gee-equation} and replace Eq.~\eqref{eq:gee-equation} by
 \begin{align}
    \left\{\i\partial_t\delta_{\underline i\, \underline k}-h^{{\rm HF},\tn{e}}_{\underline i\,\underline k}(t)\right\}G^{\tn{e}}_{\underline k\,\underline j}(t,t')
  &=
  h^{\tn{HF},\tn{es}}_{i\,\underline k}(t)G^{\tn{se}}_{k\,\underline j}(t,t')
\\
  &\quad
  +\delta_{\underline i\, \underline j}\delta_C(t,t')\,,
  \label{eq:gee-equation-new}
\end{align}
where the Green function of the environment that obeys Eq.~\eqref{eq:gee-equation-new} is now denoted by $G^{\tn{e}}$. In contrast to the former system that involved the environment Green function $g^{\tn{e}}$ that obeys Eq.~\eqref{eq:gee-equation}, the new system, obviously, conserves the total particle number. It is easily seen that, when computing the total particle number, the charge transfer terms in the equations for $G^{\tn{s}}$ and $G^{\tn{e}}$ compensate each other. We now use Eq.~\eqref{eq:gee-equation-new} and re-derive the equations for the G1--G2 embedding scheme. We proceed exactly like in Sec.~\ref{s:g1-g2}, so it is sufficient to sketch the main steps. Details of the derivation are given in Appendix~\ref{app:derivation-new}.
\begin{enumerate}
\item One easily verifies that the solutions of Eqs.~\eqref{eq:gee-equation} and \eqref{eq:gee-equation-new} are connected by
\begin{align}
 \,\,G^{\tn{e}}_{\underline l\,\underline j}(t,t')
&=g^{\tn{e}}_{\underline l\,\underline j}(t,t') + \int_C\!\!\!\d\bar{t}\,g^{\tn{e}}_{\underline l\,\underline i}(t,\bar t)  h^{\tn{HF},\tn{es}}_{\underline i\, k}(\bar t)G^{\tn{se}}_{k\,\underline j}(\bar t,t')\,.
\quad
\label{eq:connection-ge-Ge}
\end{align}
\item The solution for $G^{\rm es}$, Eq.~\eqref{eq:ges-solution}, remains unchanged.
\item The embedding selfenergy, Eq.~\eqref{eq.sigma.ct}, remains unchanged. The same applies to all two-time embedding results.
\item A crucial modification occurs upon the transition to the time-diagonal expressions of the G1--G2 scheme: The HF-GKBA has to be modified to
\begin{align}
&G^{\tn{s},\gtrless}_{ij}(t,t') =
\i \left[G_{ik}^{\tn{s},{\rm R}}(t,t') G_{kj}^{\tn{s},\gtrless}(t')-G_{ik}^{\tn{s},\gtrless}(t) G_{kj}^{\tn{s},{\rm A}}(t,t')\right]\\
&
\quad +\i \left[G_{i \,\underline k}^{\tn{se},{\rm R}}(t,t') G_{\underline k \,j}^{\tn{es},\gtrless}(t')-G_{i \,\underline k}^{\tn{se},\gtrless}(t) G_{\underline k\, j}^{\tn{es},{\rm A}}(t,t')\right]
\,,\quad
\label{eq:GKBA-ss}
\end{align}
and also includes contributions from the retarded and advanced functions that couple the system parts, $G^{\tn{se},{\rm R/A}}$.
\item With this, the time derivative of $G^{\rm es, <}(t)$ can be computed as described in Appendix.~\ref{app:derivation-new}.
\end{enumerate}
We summarize the final set of equations for the charge transfer and environment Green functions which we refer to as \textit{extended embedding scheme}:
\begin{align}
\i \frac{\d }{\d t}G^{\tn{es},<}_{\underline i\,j}(t)
&=  \left( h^{\tn{HF,es}} G^{\tn{s},<}\right)^\tn{s}_{\underline i\,j,t}
 - \left(G^{\tn{e},<}h^{\tn{HF},\tn{es}}\right)^\tn{e}_{\underline i\,j,t}
 \label{eq:ges-new}
\\
\nonumber
&\quad+ \left( h^{\tn{HF},\tn{e}}G^{\tn{es},<}\right)^\tn{e}_{\underline i\,j,t}
 - \left(G^{\tn{es},<}h^{\tn{HF},\tn{s}}\right)^\tn{s}_{\underline i\,j,t}\,,
\\
\i \frac{\tn {d}}{\tn{d}t} G^{\tn{e},<}_{\underline i\, \underline j}(t) &= \left[ h^{\rm HF,\tn{e}},G^{\tn{e},<}\right]^\tn{e}_{\underline i \,\underline j ,t}
    \label{eq:gee-new}\\
&\quad+ \left(h^{\tn{HF},\tn{es}}G^{\tn{se},<}\right)^\tn{s}_{\underline i\, \underline j,t}- \left(G^{\tn{es},<}h^{\tn{HF},\tn{se}}\right)^\tn{s}_{\underline i\, \underline j,t}\,.
\nonumber
\end{align}
Note that the equation
for $G^{\tn{es},<}$, Eq.~\eqref{eq:ges-new}, remained formally the same as before, except for the replacement $g^{\tn{e},<} \to G^{\tn{e},<}$. The main new ingredient is, of course, Eq.~\eqref{eq:gee-new} for the time evolution of the environment density matrix.

Equations~\eqref{eq:ges-new} and \eqref{eq:gee-new} are the main result of this section. They constitute the extension of the G1--G2 embedding scheme to situations of strong system-environment coupling. To verify the correctness of these equations, we apply them to the charge transfer model studied above and, in particular, to the cases that were presented in Fig.~\ref{fig:6sites}. The new results are also depicted in this figure by dashed colored lines. In all cases these lines exactly coincide with the two-time NEGF embedding results.

\subsection{Charge transfer simulations for larger systems}\label{ss:large-systems}

After verifying the correctness of the extended embedding scheme, Eqs.~\eqref{eq:time-diagonal-gless-equation}, \eqref{eq:ges-new} and \eqref{eq:gee-new}, we now take advantage of the time-linear scaling behavior of the \mbox{G1--G2} approach and apply it to significantly larger systems. We choose the same charge transfer model as introduced in Sec.~\ref{ss:g-transfer}, but consider a one-dimensional Hubbard chain of $L=50$ sites which is sufficiently long such that density reflections at the other end do not influence the charge transfer results for the parameters considered. We underline that this system is
already challenging for full
two-time NEGF simulations but, based on the comparisons presented above, we expect that our G1--G2 simulations have predictive power. Moreover, we  study the resonant charge transfer more in detail. In particular, we analyze the dependence of the charge transfer on the value of the energy $\epsilon$ of the external site.

\begin{figure}[h]
\centering
\includegraphics[width=0.4825\textwidth]{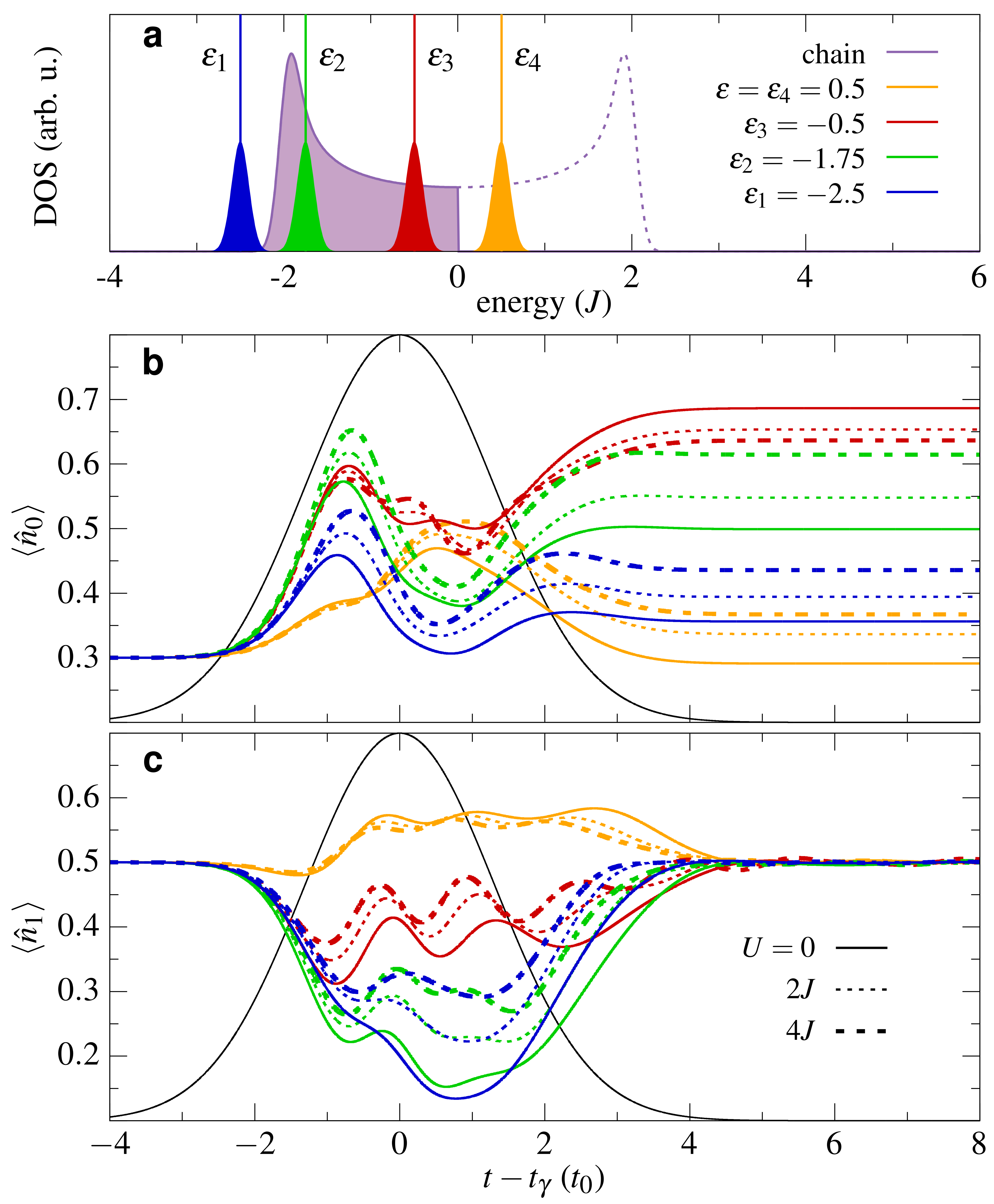}
\caption{(a) Density of states of the $50$-site Hubbard chain and four cases of the position of the energy $\epsilon$ of the additional site [for better visibility all discrete states were Gaussian-broadened]. (b) and (c): Time evolution of the density on the attached site ``0'' and on the first site ``1'' of the chain, respectively. Parameters: $\gamma_0=2J$, $n_0^{\textup{e}}=0.3$ and $\tau_\gamma=1.311t_0$. The line styles distinguish different Hubbard interaction strengths~$U$.}
\label{fig:50sites-spectrum}
\end{figure}

\begin{figure}[h]
\centering
\includegraphics[width=0.4825\textwidth]{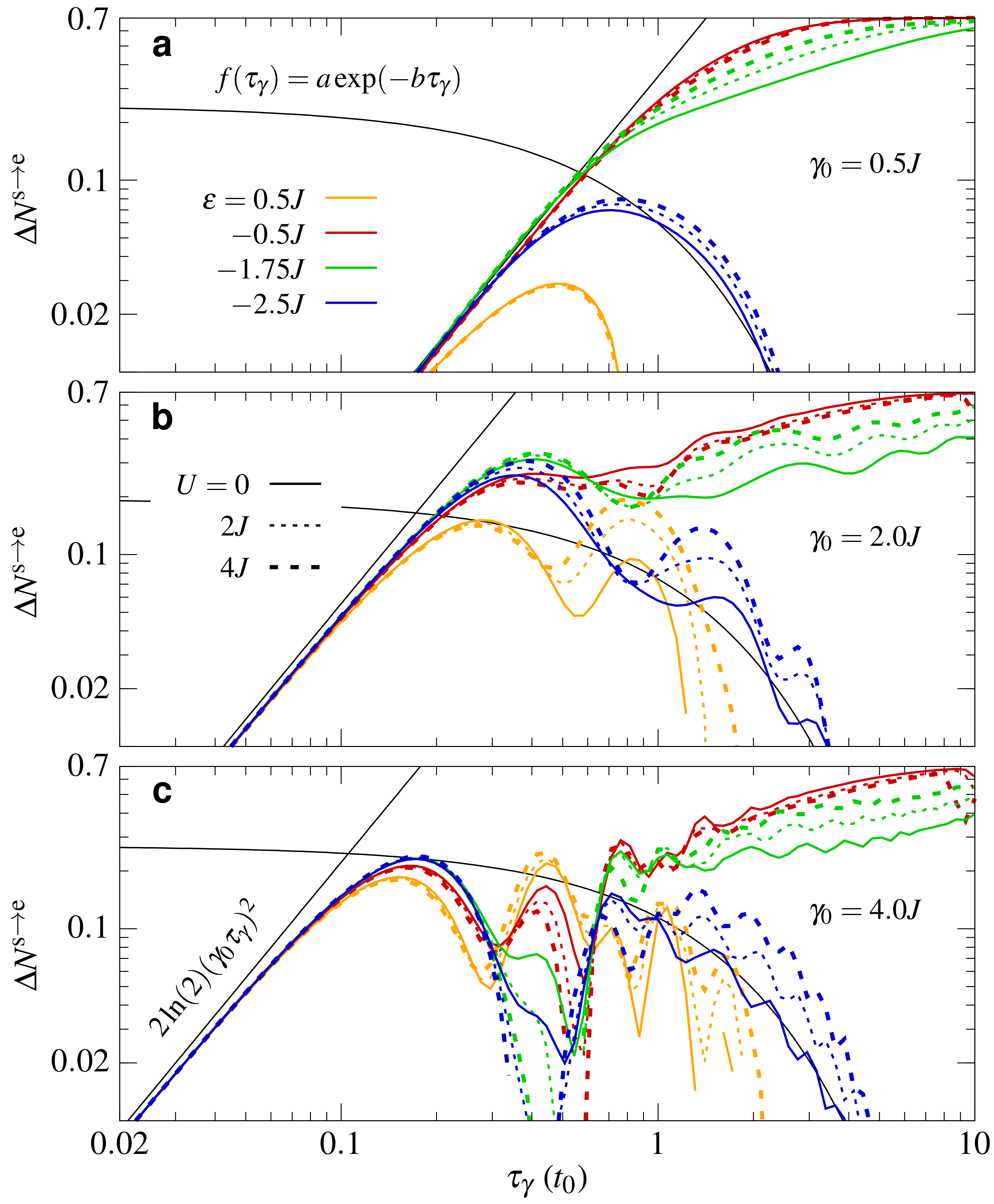}
\caption{Total charge transfer $\Delta N^{\tn{s}\rightarrow\tn{e}}$ as function of the pulse length $\tau_\gamma$ for the system of Fig.~\ref{fig:50sites-spectrum}:~(a) $\gamma_0=0.5J$, (b)~$\gamma_0=2J$, and (c)~$\gamma_0=4J$. In the numerical simulations, we used $t_\gamma=50t_0$ and extracted the value of $\Delta N^{\tn{s}\rightarrow\tn{e}}$ at time $t=150t_0$. For very large $\gamma_0$, reflections at the other end of the chain influence the adiabatic results around $\tau_\gamma\sim 10t_0$, cf.~the red lines in~(c). Furthermore, the thin black curves indicated $f(\tau_\gamma)=a \exp(-b\tau_\gamma)$ are fits to the tails of the blue solid lines for $\tau_\gamma\rightarrow 10t_0$.}
\label{fig:50sites}
\end{figure}

In the trivial case of $n^{\tn{e}}_0=0.5$ and $\epsilon=0$ no dynamics, in particular no charge transfer, will be triggered in the system, independently of the ratio $U/J$ and the form of $\gamma(t)$. In the following, we set $n_0^{\tn e}=0.3$ and consider four cases, where the energy $\epsilon$  is located either within or outside of the chain's density of states which has a bandwidth $W=4J$, cf.~Fig.~\ref{fig:50sites-spectrum}a.
Moreover, we vary the coupling parameter $U$ of the chain and the pulse parameters $\gamma_0$ and $\tau_\gamma$. As quantity of primary interest, we consider the total transferred charge from the chain to the attached site,
\begin{align}
\Delta N^{\tn{s}\rightarrow\tn{e}}&=N^{\tn s}(0)-N^{\tn s}(t\rightarrow\infty)\,,\\
N^{\tn s}(t)&=\sum_i\langle \hat{n}^{\tn{s}}_i\rangle(t)=-\i \sum_i G_{ii}^{\tn{s}, <}(t)\,,
\end{align}
as well as the densities on sites ``0'' and ``1''.

For fixed parameters, there exist different regimes which lead to characteristic responses of the system: (a)~$\tau_\gamma\gg t_0$ (adiabatic regime), (b)~$\tau_\gamma\rightarrow 0$ (perturbative regime), and (c)~$\tau_\gamma\sim t_0$ (intermediate regime). In our simulations, we have studied the full range between $\tau_\gamma=0.02t_0$ and $10 t_0$ for different values of $U$, $\gamma_0$ and $\epsilon$. In Fig.~\ref{fig:50sites-spectrum}, we concentrate on the most interesting case of the intermediate regime. The general trends are as expected: the charge transfer is strongest when the energy level $\epsilon$ is inside the Hubbard band (red and green curves) and is significantly lowered in the opposite case (blue and yellow curves). Note, that the short pulse duration plays a significant role. In contrast, for very broad pulses (slow projectiles in an ion impact scenario) we would approach Fermi's golden rule, and the overall charge transfer for the blue and yellow curves would approach zero. This is fully confirmed in Fig.~\ref{fig:50sites}, where we show the behavior for a broad range of pulse durations and three interactions strengths. Indeed, for sufficiently long pulses, the charge transfer to the additional site practically vanishes for the off-resonant cases.

In addition to the finite interaction time, also electron-electron interactions inside the chain play an important role. In the present model, the Hartree-Fock term in Eq.~\eqref{eq:hf-ham}
acts as an additional local potential. Thus, depending on the time evolution of the local density $\langle\hat{n}_1^{\tn{s}}\rangle(t)=-\i G_{11}^{\tn{s},<}(t)$, the resonance situation with the energy level $\epsilon$ may change as function of time. In the intermediate regime, this should have an essential influence on the charge transfer $\Delta N^{\tn{s}\rightarrow\tn{e}}$, particularly for larger Hubbard interactions. This is exactly what we observe: for the off-resonant cases (yellow and blue lines), increase of $U$ ``tunes'' the energy $\epsilon$ into the renormalized band, and the charge transfer increases. While a similar interaction-induced enhancement is observed also for the resonant case of $\epsilon_2$, in the second resonant case (red curve) interactions lead to a partial de-tuning of the energy $\epsilon_3$ away from the resonance, and the charge transfer is slightly reduced. The analysis of the interaction dependence is extended to a broad range of pulse durations in Fig.~\ref{fig:50sites}. For long pulses, interaction effects have the strongest influence, whereas for short pulses with $\tau_\gamma \lesssim 0.3$ interaction effects have no time to build up and practically do not affect the charge transfer.

\section{Discussion}\label{s:discussion}

In this paper, we have extended the NEGF embedding concept to the time-local HF-GKBA model -- the G1--G2 scheme. Any two-time embedding result that was reported previously, can now be translated into a time-local version. This has the benefit of time-linear scaling and the possibility of long simulation times. We have demonstrated the G1--G2 embedding scheme numerically for the example of charge transfer between an interacting Hubbard cluster and a single external site and observed excellent agreement with two-time NEGF simulations for the case of weak system-environment coupling. However, in cases of strong coupling, the time-local embedding equations drastically deviate from the two-time results, and we have shown how they can be generalized to properly account for the dynamics of the environment. An interesting observation is that our starting point -- the two-time embedding selfenergy formulation -- has a remarkable advantage: it does not require an update of the state of the environment and works for weak \textit{and} strong coupling.

Moreover, also higher-order correlation selfenergies, such as the $T$-matrix, $GW$ approximation or the dynamically screened ladder approximation~\cite{joost_prb_22} that were too costly or not accessible in two-time calculations or earlier GKBA simulations, can now be used for embedding simulations. Aside from the choice of the selfenergy, our scheme involves two approximations which we briefly discuss. The first is the choice of Hartree-Fock propagators in the GKBA (i.e. HF-GKBA). Our previous tests showed that, for finite systems, this approximation is of the same quality as two-time NEGF results, e.g. Ref.~\cite{schluenzen_prb17}, regardless of the chosen selfenergy. On the other hand, for macroscopic systems such as the electron gas or electron-hole plasmas, the HF-GKBA is not always as accurate as two-time NEGF simulations, and it may, furthermore, exhibit instabilities for long times. This was shown to be due to aliasing effects which can be mitigated using a small damping of the propagators \cite{makait_cpp_23}. A more systematic approach would use correlated propagators, as proposed in Refs.~\cite{bonitz-etal.99epjb,bonitz_qkt}. The second approximation is the use of the Hartree-Fock approximation for the environment Green function, $G^\tn{e}$, as well as for the coupling function, $G^{\tn{es}}$. While this is already a significant improvement over most previous embedding calculations [which used non-interacting Green functions], the validity, of course, depends on the strength of the interactions in the system parts. In fact, the embedding approach is not limited to the Hartree-Fock approximation, as we will show in a forthcoming paper.

Our results can be straightforwardly applied to a broad variety of embedding problems, including electronic transport in nanoscale systems, where macroscopic leads are treated as an ``environment'', e.g.~\cite{khosvari_prb_12,rabani_jcp_13}, to photoionization of atoms and molecules where the continuum states are regarded as ``environment''~\cite{perfetto_pra_15}, to the dynamics of excitonic insulators~\cite{tuovinen_prb_20}, or to the charge transfer during the impact of a  projectile onto a solid~\cite{balzer_cpp_21}.

Let us summarize the resulting time-local equations of our extended NEGF embedding scheme using the notations \eqref{eq:def-product} and \eqref{eq:def-commutator}. The equation of motion of the time-local one-particle Green function is now coupled to the equations for two auxiliary quantities -- one for the correlated part of the two-particle Green function, $\mathcal{G}(t)$, and one for the environment-system coupling single-particle Green function, $G^{\tn{es},<}(t)$,
\begin{align}
&    \i \frac{\d}{\d t} G^{<}_{ij}(t) -  \Big[ h^{\tn{HF}},G^< \Big]^\tn{s}_{ij,t}
\nonumber\\
& = \big( h^{\tn{HF,se}}G^{\tn{es},<} \big)^\tn{e}_{ij,t}
- \big( G^{\tn{se},<} h^{\tn{HF,es}} \big)^\tn{e}_{ij,t}
\\
&\quad\pm i\hbar \sum_{mnp}\left\{ w_{imnp}(t)\mathcal{G}_{npjm}(t)
- \mathcal{G}_{imnp}(t)w_{npjm}(t)\right\}\,, \nonumber
\\
&    \i \frac{\d}{\d t} \mathcal{G}_{ijkl}(t) - \Big[ h^{(2),\tn{HF}},\mathcal{G} \Big]^\tn{s}_{ijkl,t} = \Psi^\pm_{ijkl}(t)
 \, ,\quad
 \nonumber
\\
& \i \frac{\d}{\d t} G^{\tn{es},<}_{\underline \alpha\,j}(t)
- \left\{\left( h^{\tn{HF,e}} G^{\tn{es},<}\right)^\tn{e}_{\underline \alpha\, j,t} -
\left(
G^{\tn{es},<} h^{\tn{HF}}\right)^\tn{s}_{\underline \alpha \,j,t}
\right\}\\
& \qquad\quad =  \left(h^{\tn{HF,es}} G^<\right)^\tn{s}_{\underline\alpha\, j,t} - \left( G^{\tn{e},<}h^{\tn{HF,es}}\right)^\tn{e}_{\underline\alpha \,j,t}\,,
\nonumber
\\
& \i \frac{\d}{\d t} G^{\tn{e},<}_{\underline \alpha\,\underline j}(t)
- \left[ h^{\tn{HF,e}},G^{\tn{e},<}\right]^\tn{e}_{\underline\alpha\, \underline j,t} =
\\
& \qquad\quad =  \left( h^{\tn{HF,es}} G^{\tn{se},<}\right)^\tn{s}_{\underline\alpha\,\underline j,t}  - \left( G^{\tn{es},<} h^{\tn{HF,se}}\right)^\tn{e}_{\underline\alpha \,\underline j,t}\,,
\nonumber
 \end{align}
where the indices $i,j,k,l,m,n,p$ refer to the system orbitals $\phi^{\tn{s}}$, and the underlined indices $\underline \alpha, \underline j$ correspond to the environment functions $\chi^{\tn{e}}$.

It is characteristic for the G1--G2 scheme that the place of the two-time selfenergies is taken over by a set of time-local functions: $\Sigma^{\rm cor}$ gives rise to $\mathcal{G}$, whereas $\Sigma^{\rm emb}$ determines $G^{\tn{es},<}$ and $G^{\tn{e},<}$.
Note that the apparent asymmetry between the correlation and embedding selfenergies which are associated to a two-particle correlation function and single-particle Green function, respectively, is due the special embedding approximation imposed during the derivation. Going back to the Keldysh-Kadanoff-Baym equations~\eqref{eq.kbe_interface}, the interaction terms (selfenergies) originally appear in a fully symmetric way with respect to parts (s,e) of the total system. If we were to treat the environment part on the same level of accuracy as the system part, the component~\eqrefs{eq:ges-equation}{eq:gee-equation} would have the same structure as the system equation~\eqref{eq:gss-equation}, containing full correlation selfenergies. Such a symmetric treatment of all system parts has been successfully applied to multiband or multilevel systems, e.g.~\cite{bonitz_qkt}, and it has been extensively used in semiconductor optics in the frame of the semiconductor Bloch equations~\cite{lindberg_prb88}, for a two-time NEGF version of these equations, see e.g.~Refs.~\cite{kwong-etal.98pss}. In the context of a two-band system (containing e.g.~one valence and one conduction band, ``v'' and ``c'', respectively), the present coupling function $G^{\tn{es}}$ corresponds to the interband polarization function $G_{\tn{cv}}$.

In contrast, the present embedding approach aims at a simplified treatment of the environment and the system-environment coupling, on the Hartree-Fock level. The resulting equation for the coupling Green function is easily recovered from the full multi-band equations. Obviously, the neglect of the correlation selfenergy for the environment part leads to single-particle (Hartree-Fock) equations for $G^{\tn{es}}$ and  $G^{\rm e}$. They follow straightforwardly by considering the time-dependent Hartree-Fock equation for the operator Green function (we suppress the orbital indices), $G=G^{\alpha \beta}$, Eq.~\eqref{eq:g-dm},
\begin{align}
    \i\partial_t G - [h^{\rm HF},G] = 0\,,
    \nonumber
\end{align}
and by computing the ``matrix elements'', $G^{\tn{es}}=\langle \tn{e}|G|\tn{s}\rangle$ and
$G^{\tn{e}}=\langle \tn{e}|G|\tn{e}\rangle$ of this equation~\cite{bonitz_qkt}:
\begin{align}
\i\partial_t G^{\rm es} - \sum_{\beta=\tn{e},\tn{s}} \left\{h^{\rm HF, \tn{e}\beta}G^{\beta \tn{s}} - G^{ \tn{e} \beta}h^{\rm HF,\beta \tn{s}}\right\} = 0\,,
    \nonumber\\
        \i\partial_t G^{\rm e} - \sum_{\beta=\tn{e},\tn{s}} \left\{h^{\rm HF, \tn{e}\beta}G^{\beta \tn{e}} - G^{ \tn{e} \beta}h^{\rm HF,\beta \tn{e}}\right\} = 0\,.
    \nonumber
\end{align}
One readily verifies that these are exactly the operator versions of the above equations for the matrix functions $G^{\tn{ es},<}_{\underline i\, j}$ and $G^{\tn{e},<}_{\underline i\,\underline j}$, respectively. In fact, the present extended set of time-local equations that is equivalent to the NEGF embedding equations, can be derived directly from the Keldysh-Kadanoff-Baym equations without introducing an embedding selfenergy. This is shown in appendix~\ref{app:derivation-no-embedding}.

In this paper we have focused on short-time phenomena in correlated quantum systems. The equations under consideration are time-reversible. It is an interesting question for future research to connect this approach to the long-time asymptotics that are governed by irreversible equations and to
the thermodynamics and stationary transport of open systems, e.g.~\cite{haug_2008_quantum, galperin_epjst_21,spicka_long_2005, ridley_jpa_22} and references therein.

\appendix

\section{Derivation of the extended embedding scheme, equation (\ref{eq:ges-new})}\label{app:derivation-new}

For the extended embedding scheme, we start from the set of Eqs.~\eqref{eq:gss-equation}, \eqref{eq:ges-equation} and \eqref{eq:gee-equation-new} and use the modified HF-GKBA, Eq.~\eqref{eq:GKBA-ss}, to evaluate the time derivative of $G^{\tn{es}, <}(t)$ [Eq.~\eqref{eq:ges-integral},
time-diagonal element of Eq.~\eqref{eq:ges-less}]:
\begin{align}
    \i\frac{\textup{d}}{\textup{d}t}G^{\tn{es},<}_{\underline i\, j}(t) &= \i\frac{\textup{d}}{\textup{d}t}\int_{t_0}^t \d \cbar{t} \, h^{\tn{HF,es}}_{\underline k\, l}(\cbar{t})\Big[g^{\tn{e},>}_{\underline i\,\underline k}(t,\cbar{t}) G^{\tn{s},<}_{lj}(\cbar{t},t) \\
    & \qquad \qquad \quad- g^{\tn{e},<}_{\underline i\,\underline k}(t,\cbar{t}) G^{\tn{s},>}_{lj}(\cbar{t},t)\Big]
    \\
    &
    =\i\left[\frac{\d }{\d t}G^{\tn{es},<}_{\underline i\,j}(t)\right]_{\int}+\i\left[\frac{\d }{\d t}G^{\tn{es},<}_{\underline i\,j}(t)\right]_{t}\,.
\end{align}
The first term is due to differentiation of the upper integration boundary,
\begin{align}
\i \left[\frac{\d }{\d t}G^{\tn{es},<}_{\underline i\,j}(t)\right]_{\int} &=
\i \,h^{\tn{HF,es}}_{\underline k l}(t)\bigg[\!\!\!\!\underbrace{g^{\tn{e},>}_{\underline i\,\underline k}(t,t)}_{=-\i \delta_{\underline i\,\underline k}+g^{\tn{e},<}_{\underline i\,\underline k}(t,t)}\!\!\!\!G^{\tn{s},<}_{lj}(t,t)
\nonumber\\
& \qquad
- g^{\tn{e},<}_{\underline i\,\underline k}(t,t)\!\!\!\!\underbrace{G^{\tn{s},>}_{lj}(t,t)}_{=-\i \delta_{lj}+G^{\tn{s},<}_{lj}(t,t)}\!\!\!\!\bigg]\nonumber\\
&= \left(h^{\tn{HF,es}} G^{\tn{s},<}\right)^\tn{s}_{\underline i\, j,t} - \left(g^{\tn{e},<} h^{\tn{HF,es}}\right)^\tn{e}_{\underline i\, j,t} \,.
\end{align}
The second term arises from the time dependence of the integrand,
\begin{align}
&\quad
\i \left[\frac{\d}{\d t}G^{\tn{es},<}_{\underline i\,j}(t)\right]_{t}\label{eq:ges-deriv_int}\\
&\quad
=\int_{t_0}^t \d\bar t\, h^{\tn{HF},\tn{es}}_{\underline k \,l}(\bar{t})
    \bigg\{
\left[\i \partial_t g^{\tn{e},>}_{\underline i\, \underline k}(t,\bar{t})\right]G^{ \tn{s},<}_{lj}(\bar{t},t)\\
&
\qquad\qquad\qquad\qquad\qquad
+g^{\tn{e},>}_{\underline i\, \underline k}(t,\bar{t})\left[\i \partial_t G^{ \tn{s},<}_{lj}(\bar{t},t)\right]\\
&
\qquad\qquad\qquad\qquad\qquad
-\left[\i \partial_t g^{\tn{e}, <}_{\underline i\, \underline k}(t,\bar{t})\right]G^{\tn{s},>}_{lj}(\bar{t},t)
\\
&
\qquad\qquad\qquad\qquad\qquad
-g^{\tn{e}, <}_{\underline i\, \underline k}(t,\bar{t})\left[\i \partial_t G^{\tn{s},>}_{lj}(\bar{t},t)\right]\bigg\}\\
&\quad
\stackrel{\textup{Eq.~\eqref{eq:ges-integral}}}{=}\int_{t_0}^t \d\bar t\, h^{\tn{HF},\tn{es}}_{\underline k \,l}(\bar{t})
    \bigg\{
 h^{\tn{HF},\tn{e}}_{\underline i\,\underline l}(t)g^{\tn{e},>}_{\underline l\, \underline k}(t,\bar{t})G^{ \tn{s},<}_{lj}(\bar{t},t)\\
&
\qquad\qquad\qquad\qquad\qquad
- h^{\tn{HF},\tn{e}}_{\underline i\,\underline l}(t)g^{\tn{e}, <}_{\underline l\, \underline k}(t,\bar{t})G^{\tn{s},>}_{lj}(\bar{t},t)\\
&
\qquad\qquad\qquad\qquad\qquad
+g^{\tn{e},>}_{\underline i\, \underline k}(t,\bar{t})\left[\i \partial_t G^{ \tn{s},<}_{lj}(\bar{t},t)\right]
\\
&\qquad\qquad\qquad\qquad\qquad
-g^{\tn{e}, <}_{\underline i\, \underline k}(t,\bar{t})\left[\i \partial_t G^{\tn{s},>}_{lj}(\bar{t},t)\right]\bigg\}\,.
\end{align}
In the first two terms of the integral on the r.h.s., we can factor out $h^{\tn{HF},\tn{e}}(t)$ and identify the definition of $G^{\tn{es},<}(t)$. Furthermore, the partial derivatives of the two-time quantities $G^{\tn{s},<}$ and $G^{\tn{s},>}$ with respect to time can be evaluated using the following property of the HF-GKBA ansatz~\eqref{eq:GKBA-ss} (cf.~Ref.~\cite{joost_prb_20}),
\begin{align}
\left.\i\partial_t G^{\tn{s},\gtrless}_{ij}(t,t')\right|_\textup{HF-GKBA} =& - G^{\tn{s},\gtrless}_{ik}(t,t')h^{\tn{HF},\tn{s}}_{kj}(t')\\
&
- G^{\tn{se},\gtrless}_{i\,\underline k}(t,t')h^{\tn{HF},\tn{es}}_{\underline k\, j}(t')\,.
\end{align}
Thus, we obtain
\begin{align}
&\quad
\i \left[\frac{\d}{\d t}G^{\tn{es},<}_{\underline i\,j}(t)\right]_{t}
- \left(h^{\tn{HF},\tn{e}}G^{\tn{es},<}\right)^\tn{e}_{\underline i\, j,t} = \\
&\qquad
=\int_{t_0}^t \d\bar t\,h^{\tn{HF},\tn{es}}_{\underline k \,l}(\bar{t})
    \bigg\{-g^{\tn{e},>}_{\underline i\, \underline k}(t,\bar{t})G^{ \tn{s},<}_{lm}(\bar{t},t)h^{\tn{HF},\tn{s}}_{mj}(t)\\
&\qquad\qquad\qquad\qquad\qquad
+g^{\tn{e}, <}_{\underline i\, \underline k}(t,\bar{t})G^{ \tn{s},<}_{lm}(\bar{t},t)h^{\tn{HF},\tn{s}}_{mj}(t)\\
&\qquad\qquad\qquad\qquad\qquad
-g^{\tn{e},>}_{\underline i\, \underline k}(t,\bar{t}) G^{ \tn{se},<}_{l\,\underline m}(\bar{t},t)h^{\tn{HF},\tn{es}}_{\underline m\, j}(t)
\\
&\qquad\qquad\qquad\qquad\qquad
+g^{\tn{e}, <}_{\underline i\, \underline k}(t,\bar{t})G^{ \tn{se},<}_{l\,\underline m}(\bar{t},t)h^{\tn{HF},\tn{es}}_{\underline m\, j}(t)
\bigg\}
\\
&\qquad
 \stackrel{\textup{Eq.~\eqref{eq:ges-integral}}}{=}
 - \left(G^{\tn{es},<}h^{\tn{HF},\tn{s}}\right)^\tn{s}_{\underline i\, j,t}
 \\
&\qquad\qquad
+\int_{t_0}^t \d\bar t\,
    \bigg\{
-g^{\tn{e},>}_{\underline i\, \underline k}(t,\bar{t}) G^{ \tn{se},<}_{l\,\underline m}(\bar{t},t)h^{\tn{HF},\tn{es}}_{\underline m\, j}(t)
\\
&\qquad\qquad\qquad
+g^{\tn{e}, <}_{\underline i\, \underline k}(t,\bar{t})G^{ \tn{se},<}_{l\,\underline m}(\bar{t},t)h^{\tn{HF},\tn{es}}_{\underline m\, j}(t)\bigg\}h^{\tn{HF},\tn{es}}_{\underline k \,l}(\bar{t})
\\
&\qquad
=
 - \left(G^{\tn{es},<}h^{\tn{HF},\tn{s}}\right)^\tn{s}_{\underline i\,j,t}
-\left((G^{\tn{e},<}-g^{\tn{e},<})h^{\tn{HF},\tn{es}}\right)^\tn{e}_{\underline i\, j,t}\,,
\end{align}
where on the last equals sign, we have identified the difference $G^{\tn{e},<}-g^{\tn{e},<}$ from Eq.~\eqref{eq:connection-ge-Ge},
\begin{align}
G^{\tn{e},<}_{\underline i\,\underline j}(t)-g^{\tn{e},<}_{\underline i\,\underline j}(t)&=\int_{t_0}^t \d\bar t\,
\bigg\{g^{\tn{e},>}_{\underline i\, \underline k}(t,\bar{t})h^{\tn{HF},\tn{es}}_{\underline k \,l}(\bar{t})G^{ \tn{se},<}_{l\,\underline j}(\bar{t},t)\\
&\qquad
-g^{\tn{e}, <}_{\underline i\, \underline k}(t,\bar{t})h^{\tn{HF},\tn{es}}_{\underline k\, l}(\bar{t})G^{\tn{se},>}_{l\,\underline j}(\bar{t},t)\bigg\}\,.\label{eq:gee-less}
\end{align}
Collecting all terms together, we observe that the terms involving $g^{\tn{e},<}$ cancel. The final result is given by
\begin{align}
\i \frac{\d }{\d t}G^{\tn{es},<}_{\underline i\,j}(t)
=& \left(h^{\tn{HF,es}} G^{\tn{s},<}\right)^\tn{s}_{\underline i\, j,t}
 - \left(G^{\tn{es},<}h^{\tn{HF},\tn{s}}\right)^\tn{s}_{\underline i\,j,t}
\\
\nonumber
&
+\left(h^{\tn{HF},\tn{e}}G^{\tn{es},<}\right)^\tn{e}_{\underline i\, j,t}
 -\left(G^{\tn{e},<}h^{\tn{HF},\tn{es}}\right)^\tn{e}_{\underline i\, j,t}\,,
\end{align}
which is the result presented in Eq.~\eqref{eq:ges-new}.

\section{Alternative derivation of the extended embedding scheme for Hartree-Fock selfenergies}\label{app:derivation-no-embedding}

Starting from the equations~\eqref{eq:gss-equation}, \eqref{eq:ges-equation} and \eqref{eq:gee-equation-new}
on the Keldysh contour,
where we drop the collision term (putting $\Sigma^{\tn{s}} \to 0$), we take the ``$<$'' component of all equations:
\begin{align}
  \i\partial_t G^{\tn{s},<}_{ij}(t,t')-h^{{\rm HF},\tn{s}}_{ik}(t)G^{\tn{s},<}_{kj}(t,t') &=h^{{\rm HF},\tn{se}}_{i\,\underline k}(t)G^{\tn{es},<}_{\underline k \,j}(t,t') \label{eq:gssless-equation}\\
  \i\partial_t G^{\tn{es},<}_{\underline i \,j}(t,t')-h^{{\rm HF},\tn{e}}_{\underline i \,\underline k}(t)G^{\tn{es},<}_{\underline k \,j}(t,t') &=h^{{\rm HF},\tn{es}}_{\underline i\, k}(t)G^{\tn{s},<}_{kj}(t,t')\,,
\label{eq:gesless-equation}\\
  \i\partial_t G^{\tn{e},<}_{\underline i\,\underline j}(t,t')-h^{{\rm HF},\tn{e}}_{\underline i\,\underline k}(t)G^{\tn{e},<}_{\underline k\,\underline j}(t,t') &=
h^{{\rm HF},\tn{es}}_{\underline i \,k}(t)G^{\tn{se},<}_{k\,\underline j}(t,t')\,.
\label{eq:geeless-equation}
\end{align}
To derive the equations on the time diagonal, we first compute the complex adjoint of these equations and then use the symmetries,
\begin{align}
    [G^{\tn{s},<}_{kj}(t,t')]^* & = - G^{\tn{s},<}_{jk}(t',t)\,,\\
    [G^{\tn{es},<}_{\underline k \,j}(t,t')]^* & = - G^{\tn{se},<}_{j\,  \underline k}(t',t)\,,\\
    [h^{{\rm HF},\tn{es}}_{\underline i\, k}(t)]^* &= h^{{\rm HF},\tn{se}}_{k\,\underline i}(t)\,.
\end{align}
The adjoint of Eq.~\eqref{eq:geeless-equation} is given by
\begin{align}
0 &= -\i\partial_t [G^{\tn{e},<}_{\underline i\,\underline j}(t,t')]^*- [G^{\tn{e},<}_{\underline k\,\underline j}(t,t')]^* [h^{{\rm HF},\tn{e}}_{\underline i\, \underline k}(t)]^*  - ... \, \nonumber\\
&= \i\partial_t G^{\tn{e},<}_{\underline j\,\underline i}(t',t) +G^{\tn{e},<}_{\underline j\,\underline k}(t',t)h^{{\rm HF},\tn{e}}_{\underline k \,\underline i}(t) - ... \nonumber\\
&= \i\partial_{t'} G^{\tn{e},<}_{\underline i\,\underline j}(t,t') +G^{\tn{e},<}_{\underline i\,\underline k}(t,t')h^{{\rm HF},\tn{e}}_{\underline k \,\underline j}(t')\\
&\qquad
-G^{\tn{se},<}_{\underline i\, k}(t,t') h^{{\rm HF},\tn{se}}_{k\,\underline j}(t')\,,\label{eq:geeless-equation*}
\end{align}
where, in the last line, we exchanged $i \leftrightarrow j$ and $t \leftrightarrow t'$.
Now, we add Eqs.~\eqref{eq:geeless-equation} and ~\eqref{eq:geeless-equation*}, taking into account that $\partial_t + \partial_{t'}=\partial_T$, where $T=(t + t')/2$. On the time diagonal, we obtain (we also add the terms from the r.h.s.)
\begin{align}
&i\partial_t G^{\tn{e},<}_{\underline i\,\underline j}(t) - \big[h^{{\rm HF},\tn{e}},G^{\tn{e},<} \big]^\tn{e}_{\underline i\,\underline j, t} \label{eq:geless-tt}\\
&\quad
=\left(h^{\tn{HF,es}<}G^{\tn{se},<}\right)^\tn{s}_{\underline i\,\underline j, t}- \left(G^{\tn{es},<} h^{\tn{HF,se}<}\right)^\tn{s}_{\underline i\,\underline j, t}\,.\nonumber
\end{align}
In similar manner, we compute the adjoint of Eq.~\eqref{eq:gssless-equation}. The l.h.s.~of this equation is transformed exactly as before (replacing the superscript $\tn{e} \to \tn{s}$), so we concentrate on the r.h.s.:
\begin{align}
&\left[h^{{\rm HF},\tn{se}}_{i\,\underline k}(t)G^{\tn{es},<}_{\underline k \,j}(t,t')  \right]^* = [G^{\tn{es},<}_{\underline k \,j}(t,t')]^* [h^{{\rm HF},\tn{se}}_{i\,\underline k}(t)]^* \nonumber\\
&\quad
=- G^{\tn{se},<}_{j\,\underline k }(t',t) h^{{\rm HF},\tn{es}}_{\underline k\, i}(t) \to - G^{\tn{se},<}_{i\,\underline k }(t,t') h^{{\rm HF},\tn{es}}_{\underline k\,j}(t')\,,
\end{align}
where, in the last expression, we exchanged $i \leftrightarrow j$ and $t \leftrightarrow t'$.
Adding this to the r.h.s. of Eq.~\eqref{eq:gssless-equation} and taking the time-diagonal limit, we obtain
\begin{align}
   & i\partial_t G^{\tn{s},<}_{i j}(t) - \big[h^{{\rm HF},\tn{s}},G^{\tn{s},<} \big]^\tn{s}_{i j,t}
    \nonumber\\
   &\quad=
   \left(h^{{\rm HF},\tn{se}}G^{\tn{es},<}\right)^\tn{e}_{i j,t} - \left(G^{\tn{se},<} h^{{\rm HF},\tn{es}}\right)^\tn{e}_{i j,t}\,. \label{eq:gsless-tt}
\end{align}
Finally, we turn to the adjoint of Eq.~\eqref{eq:gesless-equation} and transform it:
\begin{align}
&-\i\partial_t [G^{\tn{es},<}_{\underline i\,  j}(t,t')]^*- [G^{\tn{es},<}_{\underline k\, j}(t,t')]^* [h^{{\rm HF},\tn{e}}_{\underline i \,\underline k}(t)]^*  \, \nonumber
\\
&\qquad=[G^{\tn{s},<}_{kj}(t,t')]^* [h^{{\rm HF},\tn{es}}_{\underline i\, k}(t)]^*  \nonumber
\\
&\qquad=\i\partial_t G^{\tn{se},<}_{j\,\underline i  }(t',t) + G^{\tn{se},<}_{j\, \underline k}(t',t) h^{{\rm HF},\tn{e}}_{\underline k\, \underline i}(t)  \, \nonumber
\\
&\qquad=- G^{\tn{s},<}_{jk}(t',t) h^{{\rm HF},\tn{se}}_{k\, \underline i }(t)\,.
    \nonumber
\end{align}
The final step is to again exchange $\underline i \leftrightarrow j$ and $t \leftrightarrow t'$. But, in order to obtain an equation for $G^{\tn{es} <}$ on the time diagonal, we also need to exchange $\tn{s} \leftrightarrow \tn{e}$ (in case of products, only for the outer superscripts):
\begin{align}
&\i\partial_{t'} G^{\tn{es},<}_{\underline i\,  j}(t,t') + G^{\tn{e},<}_{\underline i\, \underline k}(t,t') h^{{\rm HF},\tn{es}}_{\underline k\, j}(t')  \, \nonumber
\\
&\qquad=- G^{\tn{es},<}_{\underline i\, k}(t,t') h^{{\rm HF},\tn{s}}_{k j }(t') \,.
    \nonumber
\end{align}
Now, we again add this equation to its adjoint, Eq.~\eqref{eq:gesless-equation}, and take the time-diagonal limit:
\begin{align}
&\i\partial_{t} G^{\tn{es},<}_{\underline i\,  j}(t) -
  \left(h^{{\rm HF},\tn{e}}G^{\tn{es},<}\right)^\tn{s}_{\underline i \, j,t} + \left(G^{\tn{es},<} h^{{\rm HF},\tn{s}}\right)^\tn{e}_{\underline i \, j,t} \nonumber
\\
&\qquad=\left(h^{{\rm HF},\tn{es}}G^{\tn{s},<}\right)^\tn{s}_{\underline i \, j,t} - \left(G^{\tn{e},<} h^{{\rm HF},\tn{es}}\right)^\tn{e}_{\underline i \, j,t} \,.
    \label{eq:gesless-tt}
\end{align}
This is the final result. The three \eqrefss{eq:gsless-tt}{eq:geless-tt}{eq:gesless-tt}
exactly agree with the extended G1--G2 equations derived in the main manuscript [cf.~Sec.~\ref{s:discussion}], starting from the embedding selfenergy formulation.

\section*{Acknowledgements}
This work was supported by the Deutsche Forschungsgemeinschaft via grant BO1366/16 and high-performance computing resources of the major research instrumentation programme no.~440395346 (caucluster). This work was partly funded by the Center for Advanced Systems Understanding (CASUS) which is financed by Germany’s Federal Ministry of Education and Research (BMBF) and by the Saxon Ministry for Science, Culture and Tourism (SMWK) with tax funds on the basis of the budget approved by the Saxon State Parliament.



\begin{thebibliography}{50}%
\makeatletter
\providecommand \@ifxundefined [1]{%
 \@ifx{#1\undefined}
}%
\providecommand \@ifnum [1]{%
 \ifnum #1\expandafter \@firstoftwo
 \else \expandafter \@secondoftwo
 \fi
}%
\providecommand \@ifx [1]{%
 \ifx #1\expandafter \@firstoftwo
 \else \expandafter \@secondoftwo
 \fi
}%
\providecommand \natexlab [1]{#1}%
\providecommand \enquote  [1]{``#1''}%
\providecommand \bibnamefont  [1]{#1}%
\providecommand \bibfnamefont [1]{#1}%
\providecommand \citenamefont [1]{#1}%
\providecommand \href@noop [0]{\@secondoftwo}%
\providecommand \href [0]{\begingroup \@sanitize@url \@href}%
\providecommand \@href[1]{\@@startlink{#1}\@@href}%
\providecommand \@@href[1]{\endgroup#1\@@endlink}%
\providecommand \@sanitize@url [0]{\catcode `\\12\catcode `\$12\catcode
  `\&12\catcode `\#12\catcode `\^12\catcode `\_12\catcode `\%12\relax}%
\providecommand \@@startlink[1]{}%
\providecommand \@@endlink[0]{}%
\providecommand \url  [0]{\begingroup\@sanitize@url \@url }%
\providecommand \@url [1]{\endgroup\@href {#1}{\urlprefix }}%
\providecommand \urlprefix  [0]{URL }%
\providecommand \Eprint [0]{\href }%
\providecommand \doibase [0]{https://doi.org/}%
\providecommand \selectlanguage [0]{\@gobble}%
\providecommand \bibinfo  [0]{\@secondoftwo}%
\providecommand \bibfield  [0]{\@secondoftwo}%
\providecommand \translation [1]{[#1]}%
\providecommand \BibitemOpen [0]{}%
\providecommand \bibitemStop [0]{}%
\providecommand \bibitemNoStop [0]{.\EOS\space}%
\providecommand \EOS [0]{\spacefactor3000\relax}%
\providecommand \BibitemShut  [1]{\csname bibitem#1\endcsname}%
\let\auto@bib@innerbib\@empty
\bibitem [{\citenamefont {Xia}\ \emph {et~al.}(2015)\citenamefont {Xia},
  \citenamefont {Zundel}, \citenamefont {Carrasquilla}, \citenamefont
  {Reinhard}, \citenamefont {Wilson}, \citenamefont {Rigol},\ and\
  \citenamefont {Weiss}}]{xia_quantum_2015}%
  \BibitemOpen
  \bibfield  {author} {\bibinfo {author} {\bibfnamefont {L.}~\bibnamefont
  {Xia}}, \bibinfo {author} {\bibfnamefont {L.~A.}\ \bibnamefont {Zundel}},
  \bibinfo {author} {\bibfnamefont {J.}~\bibnamefont {Carrasquilla}}, \bibinfo
  {author} {\bibfnamefont {A.}~\bibnamefont {Reinhard}}, \bibinfo {author}
  {\bibfnamefont {J.~M.}\ \bibnamefont {Wilson}}, \bibinfo {author}
  {\bibfnamefont {M.}~\bibnamefont {Rigol}},\ and\ \bibinfo {author}
  {\bibfnamefont {D.~S.}\ \bibnamefont {Weiss}},\ }\bibfield  {title} {\bibinfo
  {title} {Quantum distillation and confinement of vacancies in a doublon
  sea},\ }\href {https://doi.org/10.1038/nphys3244} {\bibfield  {journal}
  {\bibinfo  {journal} {Nat. Phys.}\ }\textbf {\bibinfo {volume} {11}},\
  \bibinfo {pages} {316} (\bibinfo {year} {2015})}\BibitemShut {NoStop}%
\bibitem [{\citenamefont {Schl{\"u}nzen}\ \emph {et~al.}(2016)\citenamefont
  {Schl{\"u}nzen}, \citenamefont {Hermanns}, \citenamefont {Bonitz},\ and\
  \citenamefont {Verdozzi}}]{schluenzen_prb16}%
  \BibitemOpen
  \bibfield  {author} {\bibinfo {author} {\bibfnamefont {N.}~\bibnamefont
  {Schl{\"u}nzen}}, \bibinfo {author} {\bibfnamefont {S.}~\bibnamefont
  {Hermanns}}, \bibinfo {author} {\bibfnamefont {M.}~\bibnamefont {Bonitz}},\
  and\ \bibinfo {author} {\bibfnamefont {C.}~\bibnamefont {Verdozzi}},\
  }\bibfield  {title} {\bibinfo {title} {Dynamics of strongly correlated
  fermions:\textit{Ab initio} results for two and three dimensions},\ }\href
  {https://doi.org/10.1103/PhysRevB.93.035107} {\bibfield  {journal} {\bibinfo
  {journal} {Phys. Rev. B}\ }\textbf {\bibinfo {volume} {93}},\ \bibinfo
  {pages} {035107} (\bibinfo {year} {2016})}\BibitemShut {NoStop}%
\bibitem [{\citenamefont {Jensen}\ \emph {et~al.}(2013)\citenamefont {Jensen},
  \citenamefont {Ulbricht}, \citenamefont {Narita}, \citenamefont {Feng},
  \citenamefont {M\"ullen}, \citenamefont {Hertel}, \citenamefont
  {Turchinovich},\ and\ \citenamefont {Bonn}}]{jensen_ultrafast_2013}%
  \BibitemOpen
  \bibfield  {author} {\bibinfo {author} {\bibfnamefont {S.~A.}\ \bibnamefont
  {Jensen}}, \bibinfo {author} {\bibfnamefont {R.}~\bibnamefont {Ulbricht}},
  \bibinfo {author} {\bibfnamefont {A.}~\bibnamefont {Narita}}, \bibinfo
  {author} {\bibfnamefont {X.}~\bibnamefont {Feng}}, \bibinfo {author}
  {\bibfnamefont {K.}~\bibnamefont {M\"ullen}}, \bibinfo {author}
  {\bibfnamefont {T.}~\bibnamefont {Hertel}}, \bibinfo {author} {\bibfnamefont
  {D.}~\bibnamefont {Turchinovich}},\ and\ \bibinfo {author} {\bibfnamefont
  {M.}~\bibnamefont {Bonn}},\ }\bibfield  {title} {\bibinfo {title} {Ultrafast
  {Photoconductivity} of {Graphene} {Nanoribbons} and {Carbon} {Nanotubes}},\
  }\href {https://doi.org/10.1021/nl402978s} {\bibfield  {journal} {\bibinfo
  {journal} {Nano Lett.}\ }\textbf {\bibinfo {volume} {13}},\ \bibinfo {pages}
  {5925} (\bibinfo {year} {2013})}\BibitemShut {NoStop}%
\bibitem [{\citenamefont {Perfetto}\ \emph {et~al.}(2015)\citenamefont
  {Perfetto}, \citenamefont {Uimonen}, \citenamefont {van Leeuwen},\ and\
  \citenamefont {Stefanucci}}]{perfetto_pra_15}%
  \BibitemOpen
  \bibfield  {author} {\bibinfo {author} {\bibfnamefont {E.}~\bibnamefont
  {Perfetto}}, \bibinfo {author} {\bibfnamefont {A.-M.}\ \bibnamefont
  {Uimonen}}, \bibinfo {author} {\bibfnamefont {R.}~\bibnamefont {van
  Leeuwen}},\ and\ \bibinfo {author} {\bibfnamefont {G.}~\bibnamefont
  {Stefanucci}},\ }\bibfield  {title} {\bibinfo {title} {First-principles
  nonequilibrium Green's-function approach to transient photoabsorption:
  Application to atoms},\ }\href {https://doi.org/10.1103/PhysRevA.92.033419}
  {\bibfield  {journal} {\bibinfo  {journal} {Phys. Rev. A}\ }\textbf {\bibinfo
  {volume} {92}},\ \bibinfo {pages} {033419} (\bibinfo {year}
  {2015})}\BibitemShut {NoStop}%
\bibitem [{\citenamefont {Lackner}\ \emph {et~al.}(2017)\citenamefont
  {Lackner}, \citenamefont {B\ifmmode~\check{r}\else \v{r}\fi{}ezinov\'a},
  \citenamefont {Sato}, \citenamefont {Ishikawa},\ and\ \citenamefont
  {Burgd\"orfer}}]{lackner_pra_17}%
  \BibitemOpen
  \bibfield  {author} {\bibinfo {author} {\bibfnamefont {F.}~\bibnamefont
  {Lackner}}, \bibinfo {author} {\bibfnamefont {I.}~\bibnamefont
  {B\ifmmode~\check{r}\else \v{r}\fi{}ezinov\'a}}, \bibinfo {author}
  {\bibfnamefont {T.}~\bibnamefont {Sato}}, \bibinfo {author} {\bibfnamefont
  {K.~L.}\ \bibnamefont {Ishikawa}},\ and\ \bibinfo {author} {\bibfnamefont
  {J.}~\bibnamefont {Burgd\"orfer}},\ }\bibfield  {title} {\bibinfo {title}
  {High-harmonic spectra from time-dependent two-particle
  reduced-density-matrix theory},\ }\href
  {https://doi.org/10.1103/PhysRevA.95.033414} {\bibfield  {journal} {\bibinfo
  {journal} {Phys. Rev. A}\ }\textbf {\bibinfo {volume} {95}},\ \bibinfo
  {pages} {033414} (\bibinfo {year} {2017})}\BibitemShut {NoStop}%
\bibitem [{\citenamefont {Graziani}\ \emph {et~al.}(2022)\citenamefont
  {Graziani}, \citenamefont {Moldabekov}, \citenamefont {Olson},\ and\
  \citenamefont {Bonitz}}]{graziani_cpp_21}%
  \BibitemOpen
  \bibfield  {author} {\bibinfo {author} {\bibfnamefont {F.}~\bibnamefont
  {Graziani}}, \bibinfo {author} {\bibfnamefont {Z.}~\bibnamefont
  {Moldabekov}}, \bibinfo {author} {\bibfnamefont {B.}~\bibnamefont {Olson}},\
  and\ \bibinfo {author} {\bibfnamefont {M.}~\bibnamefont {Bonitz}},\
  }\bibfield  {title} {\bibinfo {title} {Shock physics in warm dense matter -- a
  quantum hydrodynamics perspective},\ }\href {https://doi.org/10.1002/ctpp.202100170} {\bibfield  {journal}
  {\bibinfo  {journal} {Contrib. Plasma Phys.}\ }\textbf {\bibinfo {volume}
  {62}},\ \bibinfo {pages} {e202100170} (\bibinfo {year} {2022})}\BibitemShut
  {NoStop}%
\bibitem [{\citenamefont {Boolakee}\ \emph {et~al.}(2022)\citenamefont
  {Boolakee}, \citenamefont {Heide}, \citenamefont
  {Garz{\'o}n-Ram{\'\inodot}rez}, \citenamefont {Weber}, \citenamefont
  {Franco},\ and\ \citenamefont {Hommelhoff}}]{hommelhoff_nature_2022}%
  \BibitemOpen
  \bibfield  {author} {\bibinfo {author} {\bibfnamefont {T.}~\bibnamefont
  {Boolakee}}, \bibinfo {author} {\bibfnamefont {C.}~\bibnamefont {Heide}},
  \bibinfo {author} {\bibfnamefont {A.}~\bibnamefont
  {Garz{\'o}n-Ram{\'\inodot}rez}}, \bibinfo {author} {\bibfnamefont {H.~B.}\
  \bibnamefont {Weber}}, \bibinfo {author} {\bibfnamefont {I.}~\bibnamefont
  {Franco}},\ and\ \bibinfo {author} {\bibfnamefont {P.}~\bibnamefont
  {Hommelhoff}},\ }\bibfield  {title} {\bibinfo {title} {Light-field control of
  real and virtual charge carriers},\ }\href
  {https://doi.org/10.1038/s41586-022-04565-9} {\bibfield  {journal} {\bibinfo
  {journal} {Nature}\ }\textbf {\bibinfo {volume} {605}},\ \bibinfo {pages}
  {251} (\bibinfo {year} {2022})}\BibitemShut {NoStop}%
\bibitem [{\citenamefont {Niggas}\ \emph {et~al.}(2022)\citenamefont {Niggas},
  \citenamefont {Schwestka}, \citenamefont {Balzer}, \citenamefont
  {Weichselbaum}, \citenamefont {Schl\"unzen}, \citenamefont {Heller},
  \citenamefont {Creutzburg}, \citenamefont {Inani}, \citenamefont {Tripathi},
  \citenamefont {Speckmann}, \citenamefont {McEvoy}, \citenamefont {Susi},
  \citenamefont {Kotakoski}, \citenamefont {Gan}, \citenamefont {George},
  \citenamefont {Turchanin}, \citenamefont {Bonitz}, \citenamefont {Aumayr},\
  and\ \citenamefont {Wilhelm}}]{niggas_prl_22}%
  \BibitemOpen
  \bibfield  {author} {\bibinfo {author} {\bibfnamefont {A.}~\bibnamefont
  {Niggas}}, \bibinfo {author} {\bibfnamefont {J.}~\bibnamefont {Schwestka}},
  \bibinfo {author} {\bibfnamefont {K.}~\bibnamefont {Balzer}}, \bibinfo
  {author} {\bibfnamefont {D.}~\bibnamefont {Weichselbaum}}, \bibinfo {author}
  {\bibfnamefont {N.}~\bibnamefont {Schl\"unzen}}, \bibinfo {author}
  {\bibfnamefont {R.}~\bibnamefont {Heller}}, \bibinfo {author} {\bibfnamefont
  {S.}~\bibnamefont {Creutzburg}}, \bibinfo {author} {\bibfnamefont
  {H.}~\bibnamefont {Inani}}, \bibinfo {author} {\bibfnamefont
  {M.}~\bibnamefont {Tripathi}}, \bibinfo {author} {\bibfnamefont
  {C.}~\bibnamefont {Speckmann}}, \bibinfo {author} {\bibfnamefont
  {N.}~\bibnamefont {McEvoy}}, \bibinfo {author} {\bibfnamefont
  {T.}~\bibnamefont {Susi}}, \bibinfo {author} {\bibfnamefont {J.}~\bibnamefont
  {Kotakoski}}, \bibinfo {author} {\bibfnamefont {Z.}~\bibnamefont {Gan}},
  \bibinfo {author} {\bibfnamefont {A.}~\bibnamefont {George}}, \bibinfo
  {author} {\bibfnamefont {A.}~\bibnamefont {Turchanin}}, \bibinfo {author}
  {\bibfnamefont {M.}~\bibnamefont {Bonitz}}, \bibinfo {author} {\bibfnamefont
  {F.}~\bibnamefont {Aumayr}},\ and\ \bibinfo {author} {\bibfnamefont {R.~A.}\
  \bibnamefont {Wilhelm}},\ }\bibfield  {title} {\bibinfo {title} {{Ion-Induced
  Surface Charge Dynamics in Freestanding Monolayers of Graphene and
  ${\mathrm{MoS}}_{2}$ Probed by the Emission of Electrons}},\ }\href
  {https://doi.org/10.1103/PhysRevLett.129.086802} {\bibfield  {journal}
  {\bibinfo  {journal} {Phys. Rev. Lett.}\ }\textbf {\bibinfo {volume} {129}},\
  \bibinfo {pages} {086802} (\bibinfo {year} {2022})}\BibitemShut {NoStop}%
\bibitem [{\citenamefont {Hochstuhl}\ \emph {et~al.}(2014)\citenamefont
  {Hochstuhl}, \citenamefont {Hinz},\ and\ \citenamefont
  {Bonitz}}]{hochstuhl_epjst_14}%
  \BibitemOpen
  \bibfield  {author} {\bibinfo {author} {\bibfnamefont {D.}~\bibnamefont
  {Hochstuhl}}, \bibinfo {author} {\bibfnamefont {C.}~\bibnamefont {Hinz}},\
  and\ \bibinfo {author} {\bibfnamefont {M.}~\bibnamefont {Bonitz}},\
  }\bibfield  {title} {\bibinfo {title} {Time-dependent multiconfiguration
  methods for the numerical simulation of photoionization processes of
  many-electron atoms},\ }\href {https://doi.org/10.1140/epjst/e2014-02092-3}
  {\bibfield  {journal} {\bibinfo  {journal} {Eur. Phys. J. Spec. Top.}\
  }\textbf {\bibinfo {volume} {223}},\ \bibinfo {pages} {177} (\bibinfo {year}
  {2014})}\BibitemShut {NoStop}%
\bibitem [{\citenamefont {Stefanucci}\ and\ \citenamefont {van
  Leeuwen}(2013)}]{stefanucci_nonequilibrium_2013}%
  \BibitemOpen
  \bibfield  {author} {\bibinfo {author} {\bibfnamefont {G.}~\bibnamefont
  {Stefanucci}}\ and\ \bibinfo {author} {\bibfnamefont {R.}~\bibnamefont {van
  Leeuwen}},\ }\href@noop {} {\emph {\bibinfo {title} {Nonequilibrium
  {{Many-Body Theory}} of {{Quantum Systems}}: {{A Modern Introduction}}}}}\
  (\bibinfo  {publisher} {{Cambridge University Press}},\ \bibinfo {address}
  {Cambridge},\ \bibinfo {year} {2013})\BibitemShut {NoStop}%
\bibitem [{\citenamefont {Balzer}\ and\ \citenamefont
  {Bonitz}(2013)}]{balzer-book}%
  \BibitemOpen
  \bibfield  {author} {\bibinfo {author} {\bibfnamefont {K.}~\bibnamefont
  {Balzer}}\ and\ \bibinfo {author} {\bibfnamefont {M.}~\bibnamefont
  {Bonitz}},\ }\href@noop {} {\emph {\bibinfo {title} {Nonequilibrium {G}reen's
  {F}unctions Approach to Inhomogeneous Systems}}}\ (\bibinfo  {publisher}
  {Springer},\ \bibinfo {address} {Berlin Heidelberg},\ \bibinfo {year}
  {2013})\BibitemShut {NoStop}%
\bibitem [{\citenamefont {Schlünzen}\ \emph
  {et~al.}(2020{\natexlab{a}})\citenamefont {Schlünzen}, \citenamefont
  {Hermanns}, \citenamefont {Scharnke},\ and\ \citenamefont
  {Bonitz}}]{schluenzen_jpcm_19}%
  \BibitemOpen
  \bibfield  {author} {\bibinfo {author} {\bibfnamefont {N.}~\bibnamefont
  {Schlünzen}}, \bibinfo {author} {\bibfnamefont {S.}~\bibnamefont
  {Hermanns}}, \bibinfo {author} {\bibfnamefont {M.}~\bibnamefont {Scharnke}},\
  and\ \bibinfo {author} {\bibfnamefont {M.}~\bibnamefont {Bonitz}},\
  }\bibfield  {title} {\bibinfo {title} {{Ultrafast dynamics of strongly
  correlated fermions -- Nonequilibrium {Green} functions and selfenergy
  approximations}},\ }\href {https://doi.org/10.1088/1361-648X/ab2d32}
  {\bibfield  {journal} {\bibinfo  {journal} {Journal of Physics: Condensed
  Matter}\ }\textbf {\bibinfo {volume} {32}},\ \bibinfo {pages} {103001}
  (\bibinfo {year} {2020}{\natexlab{a}})}\BibitemShut {NoStop}%
\bibitem [{\citenamefont {Ridley}\ \emph {et~al.}(2022)\citenamefont {Ridley},
  \citenamefont {Talarico}, \citenamefont {Karlsson}, \citenamefont {Gullo},\
  and\ \citenamefont {Tuovinen}}]{ridley_jpa_22}%
  \BibitemOpen
  \bibfield  {author} {\bibinfo {author} {\bibfnamefont {M.}~\bibnamefont
  {Ridley}}, \bibinfo {author} {\bibfnamefont {N.~W.}\ \bibnamefont
  {Talarico}}, \bibinfo {author} {\bibfnamefont {D.}~\bibnamefont {Karlsson}},
  \bibinfo {author} {\bibfnamefont {N.~L.}\ \bibnamefont {Gullo}},\ and\
  \bibinfo {author} {\bibfnamefont {R.}~\bibnamefont {Tuovinen}},\ }\bibfield
  {title} {\bibinfo {title} {A many-body approach to transport in quantum
  systems: from the transient regime to the stationary state},\ }\href
  {https://doi.org/10.1088/1751-8121/ac7119} {\bibfield  {journal} {\bibinfo
  {journal} {Journal of Physics A: Mathematical and Theoretical}\ }\textbf
  {\bibinfo {volume} {55}},\ \bibinfo {pages} {273001} (\bibinfo {year}
  {2022})}\BibitemShut {NoStop}%
\bibitem [{\citenamefont {Lipavsk\'y}\ \emph {et~al.}(1986)\citenamefont
  {Lipavsk\'y}, \citenamefont {\ifmmode \check{S}\else
  \v{S}\fi{}pi\ifmmode~\check{c}\else \v{c}\fi{}ka},\ and\ \citenamefont
  {Velick\'y}}]{lipavski_prb_86}%
  \BibitemOpen
  \bibfield  {author} {\bibinfo {author} {\bibfnamefont {P.}~\bibnamefont
  {Lipavsk\'y}}, \bibinfo {author} {\bibfnamefont {V.}~\bibnamefont {\ifmmode
  \check{S}\else \v{S}\fi{}pi\ifmmode~\check{c}\else \v{c}\fi{}ka}},\ and\
  \bibinfo {author} {\bibfnamefont {B.}~\bibnamefont {Velick\'y}},\ }\bibfield
  {title} {\bibinfo {title} {{Generalized Kadanoff-Baym ansatz for deriving
  quantum transport equations}},\ }\href
  {https://doi.org/10.1103/PhysRevB.34.6933} {\bibfield  {journal} {\bibinfo
  {journal} {Phys. Rev. B}\ }\textbf {\bibinfo {volume} {34}},\ \bibinfo
  {pages} {6933} (\bibinfo {year} {1986})}\BibitemShut {NoStop}%
\bibitem [{\citenamefont {Schlünzen}\ \emph
  {et~al.}(2020{\natexlab{b}})\citenamefont {Schlünzen}, \citenamefont
  {Joost},\ and\ \citenamefont {Bonitz}}]{schluenzen_prl_20}%
  \BibitemOpen
  \bibfield  {author} {\bibinfo {author} {\bibfnamefont {N.}~\bibnamefont
  {Schlünzen}}, \bibinfo {author} {\bibfnamefont {J.-P.}\ \bibnamefont
  {Joost}},\ and\ \bibinfo {author} {\bibfnamefont {M.}~\bibnamefont
  {Bonitz}},\ }\bibfield  {title} {\bibinfo {title} {{Achieving the Scaling
  Limit for Nonequilibrium Green Functions Simulations}},\ }\href
  {https://doi.org/10.1103/PhysRevLett.124.076601} {\bibfield  {journal}
  {\bibinfo  {journal} {Phys. Rev. Lett.}\ }\textbf {\bibinfo {volume} {124}},\
  \bibinfo {pages} {076601} (\bibinfo {year} {2020}{\natexlab{b}})}\BibitemShut
  {NoStop}%
\bibitem [{\citenamefont {Joost}\ \emph {et~al.}(2020)\citenamefont {Joost},
  \citenamefont {Schl\"unzen},\ and\ \citenamefont {Bonitz}}]{joost_prb_20}%
  \BibitemOpen
  \bibfield  {author} {\bibinfo {author} {\bibfnamefont {J.-P.}\ \bibnamefont
  {Joost}}, \bibinfo {author} {\bibfnamefont {N.}~\bibnamefont {Schl\"unzen}},\
  and\ \bibinfo {author} {\bibfnamefont {M.}~\bibnamefont {Bonitz}},\
  }\bibfield  {title} {\bibinfo {title} {{G1-G2 scheme: Dramatic acceleration
  of nonequilibrium Green functions simulations within the Hartree-Fock
  generalized Kadanoff-Baym ansatz}},\ }\href
  {https://doi.org/10.1103/PhysRevB.101.245101} {\bibfield  {journal} {\bibinfo
   {journal} {Phys. Rev. B}\ }\textbf {\bibinfo {volume} {101}},\ \bibinfo
  {pages} {245101} (\bibinfo {year} {2020})}\BibitemShut {NoStop}%
\bibitem [{\citenamefont {Pavlyukh}\ \emph {et~al.}(2021)\citenamefont
  {Pavlyukh}, \citenamefont {Perfetto},\ and\ \citenamefont
  {Stefanucci}}]{pavlyukh_prb_21}%
  \BibitemOpen
  \bibfield  {author} {\bibinfo {author} {\bibfnamefont {Y.}~\bibnamefont
  {Pavlyukh}}, \bibinfo {author} {\bibfnamefont {E.}~\bibnamefont {Perfetto}},\
  and\ \bibinfo {author} {\bibfnamefont {G.}~\bibnamefont {Stefanucci}},\
  }\bibfield  {title} {\bibinfo {title} {{Photoinduced dynamics of organic
  molecules using nonequilibrium Green's functions with second-Born, $GW$,
  $T$-matrix, and three-particle correlations}},\ }\href
  {https://doi.org/10.1103/PhysRevB.104.035124} {\bibfield  {journal} {\bibinfo
   {journal} {Phys. Rev. B}\ }\textbf {\bibinfo {volume} {104}},\ \bibinfo
  {pages} {035124} (\bibinfo {year} {2021})}\BibitemShut {NoStop}%
\bibitem [{\citenamefont {Karlsson}\ \emph {et~al.}(2021)\citenamefont
  {Karlsson}, \citenamefont {van Leeuwen}, \citenamefont {Pavlyukh},
  \citenamefont {Perfetto},\ and\ \citenamefont {Stefanucci}}]{karlsson_prl21}%
  \BibitemOpen
  \bibfield  {author} {\bibinfo {author} {\bibfnamefont {D.}~\bibnamefont
  {Karlsson}}, \bibinfo {author} {\bibfnamefont {R.}~\bibnamefont {van
  Leeuwen}}, \bibinfo {author} {\bibfnamefont {Y.}~\bibnamefont {Pavlyukh}},
  \bibinfo {author} {\bibfnamefont {E.}~\bibnamefont {Perfetto}},\ and\
  \bibinfo {author} {\bibfnamefont {G.}~\bibnamefont {Stefanucci}},\ }\bibfield
   {title} {\bibinfo {title} {{Fast Green's Function Method for Ultrafast
  Electron-Boson Dynamics}},\ }\href
  {https://doi.org/10.1103/PhysRevLett.127.036402} {\bibfield  {journal}
  {\bibinfo  {journal} {Phys. Rev. Lett.}\ }\textbf {\bibinfo {volume} {127}},\
  \bibinfo {pages} {036402} (\bibinfo {year} {2021})}\BibitemShut {NoStop}%
\bibitem [{\citenamefont {Perfetto}\ \emph {et~al.}(2022)\citenamefont
  {Perfetto}, \citenamefont {Pavlyukh},\ and\ \citenamefont
  {Stefanucci}}]{perfetto_prl_22}%
  \BibitemOpen
  \bibfield  {author} {\bibinfo {author} {\bibfnamefont {E.}~\bibnamefont
  {Perfetto}}, \bibinfo {author} {\bibfnamefont {Y.}~\bibnamefont {Pavlyukh}},\
  and\ \bibinfo {author} {\bibfnamefont {G.}~\bibnamefont {Stefanucci}},\
  }\bibfield  {title} {\bibinfo {title} {{Real-Time $GW$: Toward an Ab Initio
  Description of the Ultrafast Carrier and Exciton Dynamics in Two-Dimensional
  Materials}},\ }\href {https://doi.org/10.1103/PhysRevLett.128.016801}
  {\bibfield  {journal} {\bibinfo  {journal} {Phys. Rev. Lett.}\ }\textbf
  {\bibinfo {volume} {128}},\ \bibinfo {pages} {016801} (\bibinfo {year}
  {2022})}\BibitemShut {NoStop}%
\bibitem [{\citenamefont {Wu}\ \emph {et~al.}(2018)\citenamefont {Wu},
  \citenamefont {Lovorn}, \citenamefont {Tutuc},\ and\ \citenamefont
  {MacDonald}}]{wu_prl_18}%
  \BibitemOpen
  \bibfield  {author} {\bibinfo {author} {\bibfnamefont {F.}~\bibnamefont
  {Wu}}, \bibinfo {author} {\bibfnamefont {T.}~\bibnamefont {Lovorn}}, \bibinfo
  {author} {\bibfnamefont {E.}~\bibnamefont {Tutuc}},\ and\ \bibinfo {author}
  {\bibfnamefont {A.~H.}\ \bibnamefont {MacDonald}},\ }\bibfield  {title}
  {\bibinfo {title} {Hubbard model physics in transition metal dichalcogenide
  moir\'e bands},\ }\href {https://doi.org/10.1103/PhysRevLett.121.026402}
  {\bibfield  {journal} {\bibinfo  {journal} {Phys. Rev. Lett.}\ }\textbf
  {\bibinfo {volume} {121}},\ \bibinfo {pages} {026402} (\bibinfo {year}
  {2018})}\BibitemShut {NoStop}%
\bibitem [{\citenamefont {Li}\ \emph {et~al.}(2021)\citenamefont {Li},
  \citenamefont {Li}, \citenamefont {Regan}, \citenamefont {Wang},
  \citenamefont {Zhao}, \citenamefont {Kahn}, \citenamefont {Yumigeta},
  \citenamefont {Blei}, \citenamefont {Taniguchi}, \citenamefont {Watanabe},
  \citenamefont {Tongay}, \citenamefont {Zettl}, \citenamefont {Crommie},\ and\
  \citenamefont {Wang}}]{li_nat_21}%
  \BibitemOpen
  \bibfield  {author} {\bibinfo {author} {\bibfnamefont {H.}~\bibnamefont
  {Li}}, \bibinfo {author} {\bibfnamefont {S.}~\bibnamefont {Li}}, \bibinfo
  {author} {\bibfnamefont {E.~C.}\ \bibnamefont {Regan}}, \bibinfo {author}
  {\bibfnamefont {D.}~\bibnamefont {Wang}}, \bibinfo {author} {\bibfnamefont
  {W.}~\bibnamefont {Zhao}}, \bibinfo {author} {\bibfnamefont {S.}~\bibnamefont
  {Kahn}}, \bibinfo {author} {\bibfnamefont {K.}~\bibnamefont {Yumigeta}},
  \bibinfo {author} {\bibfnamefont {M.}~\bibnamefont {Blei}}, \bibinfo {author}
  {\bibfnamefont {T.}~\bibnamefont {Taniguchi}}, \bibinfo {author}
  {\bibfnamefont {K.}~\bibnamefont {Watanabe}}, \bibinfo {author}
  {\bibfnamefont {S.}~\bibnamefont {Tongay}}, \bibinfo {author} {\bibfnamefont
  {A.}~\bibnamefont {Zettl}}, \bibinfo {author} {\bibfnamefont {M.~F.}\
  \bibnamefont {Crommie}},\ and\ \bibinfo {author} {\bibfnamefont
  {F.}~\bibnamefont {Wang}},\ }\bibfield  {title} {\bibinfo {title} {{Imaging
  two-dimensional generalized Wigner crystals}},\ }\href
  {https://doi.org/10.1038/s41586-021-03874-9} {\bibfield  {journal} {\bibinfo
  {journal} {Nature}\ }\textbf {\bibinfo {volume} {597}},\ \bibinfo {pages}
  {650} (\bibinfo {year} {2021})}\BibitemShut {NoStop}%
\bibitem [{\citenamefont {Smoleński}\ \emph {et~al.}(2021)\citenamefont
  {Smoleński}, \citenamefont {Dolgirev}, \citenamefont {Kuhlenkamp},
  \citenamefont {Popert}, \citenamefont {Shimazaki}, \citenamefont {Back},
  \citenamefont {Lu}, \citenamefont {Kroner}, \citenamefont {Watanabe},
  \citenamefont {Taniguchi}, \citenamefont {Esterlis}, \citenamefont {Demler},\
  and\ \citenamefont {Imamoğlu}}]{smolenski_nat_21}%
  \BibitemOpen
  \bibfield  {author} {\bibinfo {author} {\bibfnamefont {T.}~\bibnamefont
  {Smoleński}}, \bibinfo {author} {\bibfnamefont {P.~E.}\ \bibnamefont
  {Dolgirev}}, \bibinfo {author} {\bibfnamefont {C.}~\bibnamefont
  {Kuhlenkamp}}, \bibinfo {author} {\bibfnamefont {A.}~\bibnamefont {Popert}},
  \bibinfo {author} {\bibfnamefont {Y.}~\bibnamefont {Shimazaki}}, \bibinfo
  {author} {\bibfnamefont {P.}~\bibnamefont {Back}}, \bibinfo {author}
  {\bibfnamefont {X.}~\bibnamefont {Lu}}, \bibinfo {author} {\bibfnamefont
  {M.}~\bibnamefont {Kroner}}, \bibinfo {author} {\bibfnamefont
  {K.}~\bibnamefont {Watanabe}}, \bibinfo {author} {\bibfnamefont
  {T.}~\bibnamefont {Taniguchi}}, \bibinfo {author} {\bibfnamefont
  {I.}~\bibnamefont {Esterlis}}, \bibinfo {author} {\bibfnamefont
  {E.}~\bibnamefont {Demler}},\ and\ \bibinfo {author} {\bibfnamefont
  {A.}~\bibnamefont {Imamoğlu}},\ }\bibfield  {title} {\bibinfo {title}
  {{Signatures of Wigner crystal of electrons in a monolayer semiconductor}},\
  }\href {https://doi.org/10.1038/s41586-021-03590-4} {\bibfield  {journal}
  {\bibinfo  {journal} {Nature}\ }\textbf {\bibinfo {volume} {595}},\ \bibinfo
  {pages} {53} (\bibinfo {year} {2021})}\BibitemShut {NoStop}%
\bibitem [{\citenamefont {Bonitz}\ and\ \citenamefont
  {Joost}(2021)}]{bonitz_pj_21}%
  \BibitemOpen
  \bibfield  {author} {\bibinfo {author} {\bibfnamefont {M.}~\bibnamefont
  {Bonitz}}\ and\ \bibinfo {author} {\bibfnamefont {J.-P.}\ \bibnamefont
  {Joost}},\ }\bibfield  {title} {\bibinfo {title} {Wigner crystal in
  two-dimensional solids? (in german)},\ }\href@noop {} {\bibfield  {journal}
  {\bibinfo  {journal} {Physik Journal}\ }\textbf {\bibinfo {volume} {20}},\
  \bibinfo {pages} {11} (\bibinfo {year} {2021})}\BibitemShut {NoStop}%
\bibitem [{\citenamefont {Joost}\ \emph {et~al.}(2022)\citenamefont {Joost},
  \citenamefont {Schl{\"u}nzen}, \citenamefont {Ohldag}, \citenamefont {Bonitz},
  \citenamefont {Lackner},\ and\ \citenamefont {Brezinova}}]{joost_prb_22}%
  \BibitemOpen
  \bibfield  {author} {\bibinfo {author} {\bibfnamefont {J.-P.}\ \bibnamefont
  {Joost}}, \bibinfo {author} {\bibfnamefont {N.}~\bibnamefont {Schl{\"u}nzen}},
  \bibinfo {author} {\bibfnamefont {H.}~\bibnamefont {Ohldag}}, \bibinfo
  {author} {\bibfnamefont {M.}~\bibnamefont {Bonitz}}, \bibinfo {author}
  {\bibfnamefont {F.}~\bibnamefont {Lackner}},\ and\ \bibinfo {author}
  {\bibfnamefont {I.}~\bibnamefont {Brezinova}},\ }\bibfield  {title} {\bibinfo
  {title} {The dynamically screened ladder approximation: Simultaneous
  treatment of strong electronic correlations and dynamical screening out of
  equilibrium},\ }\href {https://doi.org/10.1103/PhysRevB.105.165155}
  {\bibfield  {journal} {\bibinfo  {journal} {Physical Review B}\ }\textbf
  {\bibinfo {volume} {105}},\ \bibinfo {pages} {165155} (\bibinfo {year}
  {2022})}\BibitemShut {NoStop}%
\bibitem [{\citenamefont {Schroedter}\ \emph {et~al.}(2022)\citenamefont
  {Schroedter}, \citenamefont {Joost},\ and\ \citenamefont
  {Bonitz}}]{schroedter_cmp_22}%
  \BibitemOpen
  \bibfield  {author} {\bibinfo {author} {\bibfnamefont {E.}~\bibnamefont
  {Schroedter}}, \bibinfo {author} {\bibfnamefont {J.-P.}\ \bibnamefont
  {Joost}},\ and\ \bibinfo {author} {\bibfnamefont {M.}~\bibnamefont
  {Bonitz}},\ }\bibfield  {title} {\bibinfo {title} {{Quantum Fluctuations
  Approach to the Nonequilibrium $GW$-Approximation}},\ }\href
  {https://doi.org/10.5488/CMP.25.23401} {\bibfield  {journal} {\bibinfo
  {journal} {Cond. Matt. Phys.}\ }\textbf {\bibinfo {volume} {25}},\ \bibinfo
  {pages} {23401} (\bibinfo {year} {2022})}\BibitemShut {NoStop}%
\bibitem [{\citenamefont {Hochstuhl}\ and\ \citenamefont
  {Bonitz}(2012)}]{hochstuhl_pra_12}%
  \BibitemOpen
  \bibfield  {author} {\bibinfo {author} {\bibfnamefont {D.}~\bibnamefont
  {Hochstuhl}}\ and\ \bibinfo {author} {\bibfnamefont {M.}~\bibnamefont
  {Bonitz}},\ }\bibfield  {title} {\bibinfo {title} {Time-dependent
  restricted-active-space configuration-interaction method for the
  photoionization of many-electron atoms},\ }\href
  {https://doi.org/10.1103/PhysRevA.86.053424} {\bibfield  {journal} {\bibinfo
  {journal} {Phys. Rev. A}\ }\textbf {\bibinfo {volume} {86}},\ \bibinfo
  {pages} {053424} (\bibinfo {year} {2012})}\BibitemShut {NoStop}%
\bibitem [{\citenamefont {Warshel}\ and\ \citenamefont
  {Levitt}(1976)}]{WARSHEL1976227}%
  \BibitemOpen
  \bibfield  {author} {\bibinfo {author} {\bibfnamefont {A.}~\bibnamefont
  {Warshel}}\ and\ \bibinfo {author} {\bibfnamefont {M.}~\bibnamefont
  {Levitt}},\ }\bibfield  {title} {\bibinfo {title} {Theoretical studies of
  enzymic reactions: Dielectric, electrostatic and steric stabilization of the
  carbonium ion in the reaction of lysozyme},\ }\href
  {https://doi.org/https://doi.org/10.1016/0022-2836(76)90311-9} {\bibfield
  {journal} {\bibinfo  {journal} {Journal of Molecular Biology}\ }\textbf
  {\bibinfo {volume} {103}},\ \bibinfo {pages} {227} (\bibinfo {year}
  {1976})}\BibitemShut {NoStop}%
\bibitem [{\citenamefont {{Chibani, Wael}}(2016)}]{wael-chibani-phd_16}%
  \BibitemOpen
  \bibfield  {author} {\bibinfo {author} {\bibnamefont {{Chibani, Wael}}},\
  }\emph {\bibinfo {title} {{Self-Consistent Green’s Function Embedding for
  Advanced Electronic Structure Calculations based on a Dynamical Mean-Field
  Concept}}},\ \href@noop {} {Ph.D. thesis},\ \bibinfo  {school} {{Technical
  University Berlin}}, \bibinfo {address} {{Berlin}} (\bibinfo {year}
  {2016})\BibitemShut {NoStop}%
\bibitem [{\citenamefont {Ness}(2014)}]{ness_pre_14}%
  \BibitemOpen
  \bibfield  {author} {\bibinfo {author} {\bibfnamefont {H.}~\bibnamefont
  {Ness}},\ }\bibfield  {title} {\bibinfo {title} {Nonequilibrium density
  matrix in quantum open systems: Generalization for simultaneous heat and
  charge steady-state transport},\ }\href
  {https://doi.org/10.1103/PhysRevE.90.062119} {\bibfield  {journal} {\bibinfo
  {journal} {Phys. Rev. E}\ }\textbf {\bibinfo {volume} {90}},\ \bibinfo
  {pages} {062119} (\bibinfo {year} {2014})}\BibitemShut {NoStop}%
\bibitem [{\citenamefont {Bonitz}\ \emph {et~al.}(2019)\citenamefont {Bonitz},
  \citenamefont {Filinov}, \citenamefont {Abraham}, \citenamefont {Balzer},
  \citenamefont {K{\"a}hlert}, \citenamefont {Pehlke}, \citenamefont {Bronold},
  \citenamefont {Pamperin}, \citenamefont {Becker}, \citenamefont {Loffhagen},\
  and\ \citenamefont {Fehske}}]{Bonitz_fcse_19}%
  \BibitemOpen
  \bibfield  {author} {\bibinfo {author} {\bibfnamefont {M.}~\bibnamefont
  {Bonitz}}, \bibinfo {author} {\bibfnamefont {A.}~\bibnamefont {Filinov}},
  \bibinfo {author} {\bibfnamefont {J.-W.}\ \bibnamefont {Abraham}}, \bibinfo
  {author} {\bibfnamefont {K.}~\bibnamefont {Balzer}}, \bibinfo {author}
  {\bibfnamefont {H.}~\bibnamefont {K{\"a}hlert}}, \bibinfo {author}
  {\bibfnamefont {E.}~\bibnamefont {Pehlke}}, \bibinfo {author} {\bibfnamefont
  {F.~X.}\ \bibnamefont {Bronold}}, \bibinfo {author} {\bibfnamefont
  {M.}~\bibnamefont {Pamperin}}, \bibinfo {author} {\bibfnamefont
  {M.}~\bibnamefont {Becker}}, \bibinfo {author} {\bibfnamefont
  {D.}~\bibnamefont {Loffhagen}},\ and\ \bibinfo {author} {\bibfnamefont
  {H.}~\bibnamefont {Fehske}},\ }\bibfield  {title} {\bibinfo {title} {Towards
  an integrated modeling of the plasma-solid interface},\ }\href
  {https://doi.org/10.1007/s11705-019-1793-4} {\bibfield  {journal} {\bibinfo
  {journal} {Frontiers of Chemical Science and Engineering}\ }\textbf {\bibinfo
  {volume} {13}},\ \bibinfo {pages} {201} (\bibinfo {year} {2019})}\BibitemShut
  {NoStop}%
\bibitem [{\citenamefont {Bronold}\ and\ \citenamefont
  {Fehske}(2022)}]{bronold_jap_22}%
  \BibitemOpen
  \bibfield  {author} {\bibinfo {author} {\bibfnamefont {F.~X.}\ \bibnamefont
  {Bronold}}\ and\ \bibinfo {author} {\bibfnamefont {H.}~\bibnamefont
  {Fehske}},\ }\bibfield  {title} {\bibinfo {title} {Invariant embedding
  approach to secondary electron emission from metals},\ }\href
  {https://doi.org/10.1063/5.0082468} {\bibfield  {journal} {\bibinfo
  {journal} {Journal of Applied Physics}\ }\textbf {\bibinfo {volume} {131}},\
  \bibinfo {pages} {113302} (\bibinfo {year} {2022})},\ \BibitemShut {NoStop}%
\bibitem [{\citenamefont {Khosravi}\ \emph {et~al.}(2012)\citenamefont
  {Khosravi}, \citenamefont {Uimonen}, \citenamefont {Stan}, \citenamefont
  {Stefanucci}, \citenamefont {Kurth}, \citenamefont {van Leeuwen},\ and\
  \citenamefont {Gross}}]{khosvari_prb_12}%
  \BibitemOpen
  \bibfield  {author} {\bibinfo {author} {\bibfnamefont {E.}~\bibnamefont
  {Khosravi}}, \bibinfo {author} {\bibfnamefont {A.-M.}\ \bibnamefont
  {Uimonen}}, \bibinfo {author} {\bibfnamefont {A.}~\bibnamefont {Stan}},
  \bibinfo {author} {\bibfnamefont {G.}~\bibnamefont {Stefanucci}}, \bibinfo
  {author} {\bibfnamefont {S.}~\bibnamefont {Kurth}}, \bibinfo {author}
  {\bibfnamefont {R.}~\bibnamefont {van Leeuwen}},\ and\ \bibinfo {author}
  {\bibfnamefont {E.~K.~U.}\ \bibnamefont {Gross}},\ }\bibfield  {title}
  {\bibinfo {title} {Correlation effects in bistability at the nanoscale:
  Steady state and beyond},\ }\href
  {https://doi.org/10.1103/PhysRevB.85.075103} {\bibfield  {journal} {\bibinfo
  {journal} {Phys. Rev. B}\ }\textbf {\bibinfo {volume} {85}},\ \bibinfo
  {pages} {075103} (\bibinfo {year} {2012})}\BibitemShut {NoStop}%
\bibitem [{\citenamefont {Levy}\ and\ \citenamefont
  {Rabani}(2013)}]{rabani_jcp_13}%
  \BibitemOpen
  \bibfield  {author} {\bibinfo {author} {\bibfnamefont {T.~J.}\ \bibnamefont
  {Levy}}\ and\ \bibinfo {author} {\bibfnamefont {E.}~\bibnamefont {Rabani}},\
  }\bibfield  {title} {\bibinfo {title} {Steady state conductance in a double
  quantum dot array: The nonequilibrium equation-of-motion Green function
  approach},\ }\href {https://doi.org/10.1063/1.4802752} {\bibfield  {journal}
  {\bibinfo  {journal} {The Journal of Chemical Physics}\ }\textbf {\bibinfo
  {volume} {138}},\ \bibinfo {pages} {164125} (\bibinfo {year} {2013})},\
 \BibitemShut {NoStop}%
\bibitem [{\citenamefont {Tuovinen}\ \emph {et~al.}(2020)\citenamefont
  {Tuovinen}, \citenamefont {Gole\ifmmode~\check{z}\else \v{z}\fi{}},
  \citenamefont {Eckstein},\ and\ \citenamefont {Sentef}}]{tuovinen_prb_20}%
  \BibitemOpen
  \bibfield  {author} {\bibinfo {author} {\bibfnamefont {R.}~\bibnamefont
  {Tuovinen}}, \bibinfo {author} {\bibfnamefont {D.}~\bibnamefont
  {Gole\ifmmode~\check{z}\else \v{z}\fi{}}}, \bibinfo {author} {\bibfnamefont
  {M.}~\bibnamefont {Eckstein}},\ and\ \bibinfo {author} {\bibfnamefont
  {M.~A.}\ \bibnamefont {Sentef}},\ }\bibfield  {title} {\bibinfo {title}
  {Comparing the generalized Kadanoff-Baym ansatz with the full Kadanoff-Baym
  equations for an excitonic insulator out of equilibrium},\ }\href
  {https://doi.org/10.1103/PhysRevB.102.115157} {\bibfield  {journal} {\bibinfo
   {journal} {Phys. Rev. B}\ }\textbf {\bibinfo {volume} {102}},\ \bibinfo
  {pages} {115157} (\bibinfo {year} {2020})}\BibitemShut {NoStop}%
\bibitem [{\citenamefont {Covito}\ \emph {et~al.}(2018)\citenamefont {Covito},
  \citenamefont {Perfetto}, \citenamefont {Rubio},\ and\ \citenamefont
  {Stefanucci}}]{covito_pra_18}%
  \BibitemOpen
  \bibfield  {author} {\bibinfo {author} {\bibfnamefont {F.}~\bibnamefont
  {Covito}}, \bibinfo {author} {\bibfnamefont {E.}~\bibnamefont {Perfetto}},
  \bibinfo {author} {\bibfnamefont {A.}~\bibnamefont {Rubio}},\ and\ \bibinfo
  {author} {\bibfnamefont {G.}~\bibnamefont {Stefanucci}},\ }\bibfield  {title}
  {\bibinfo {title} {Real-time dynamics of auger wave packets and decays in
  ultrafast charge migration processes},\ }\href
  {https://doi.org/10.1103/PhysRevA.97.061401} {\bibfield  {journal} {\bibinfo
  {journal} {Phys. Rev. A}\ }\textbf {\bibinfo {volume} {97}},\ \bibinfo
  {pages} {061401} (\bibinfo {year} {2018})}\BibitemShut {NoStop}%
\bibitem [{\citenamefont {Balzer}\ and\ \citenamefont
  {Bonitz}(2021)}]{balzer_cpp_21}%
  \BibitemOpen
  \bibfield  {author} {\bibinfo {author} {\bibfnamefont {K.}~\bibnamefont
  {Balzer}}\ and\ \bibinfo {author} {\bibfnamefont {M.}~\bibnamefont
  {Bonitz}},\ }\bibfield  {title} {\bibinfo {title} {Neutralization dynamics of
  slow highly charged ions passing through graphene nanoflakes -- an embedding
  self-energy approach},\ }\href {https://doi.org/10.1002/ctpp.202100041}
  {\bibfield  {journal} {\bibinfo  {journal} {Contrib. Plasma Phys.}\ }\textbf
  {\bibinfo {volume} {61}},\ \bibinfo {pages} {e202100040} (\bibinfo {year}
  {2021})}\BibitemShut {NoStop}%
\bibitem [{\citenamefont {Kadanoff}\ and\ \citenamefont
  {Baym}(1962)}]{kadanoff-baym-book}%
  \BibitemOpen
  \bibfield  {author} {\bibinfo {author} {\bibfnamefont {L.}~\bibnamefont
  {Kadanoff}}\ and\ \bibinfo {author} {\bibfnamefont {G.}~\bibnamefont
  {Baym}},\ }\href@noop {} {\emph {\bibinfo {title} {Quantum Statistical
  Mechanics}}}\ (\bibinfo  {publisher} {Benjamin},\ \bibinfo {address} {New
  York},\ \bibinfo {year} {1962})\BibitemShut {NoStop}%
\bibitem [{\citenamefont {Schl{\"u}nzen}\ and\ \citenamefont
  {Bonitz}(2016)}]{schluenzen_cpp16}%
  \BibitemOpen
  \bibfield  {author} {\bibinfo {author} {\bibfnamefont {N.}~\bibnamefont
  {Schl{\"u}nzen}}\ and\ \bibinfo {author} {\bibfnamefont {M.}~\bibnamefont
  {Bonitz}},\ }\bibfield  {title} {\bibinfo {title} {{Nonequilibrium Green
  Functions Approach to Strongly Correlated Fermions in Lattice Systems}},\
  }\href {https://doi.org/10.1002/ctpp.201610003} {\bibfield  {journal}
  {\bibinfo  {journal} {Contrib. Plasma Phys.}\ }\textbf {\bibinfo {volume}
  {56}},\ \bibinfo {pages} {5} (\bibinfo {year} {2016})}\BibitemShut {NoStop}%
\bibitem [{\citenamefont {Bonitz}(2016)}]{bonitz_qkt}%
  \BibitemOpen
  \bibfield  {author} {\bibinfo {author} {\bibfnamefont {M.}~\bibnamefont
  {Bonitz}},\ }\href@noop {} {\emph {\bibinfo {title} {{Quantum Kinetic
  Theory}}}},\ \bibinfo {edition} {2nd}\ ed.,\ Teubner-Texte zur Physik\
  (\bibinfo  {publisher} {Springer},\ \bibinfo {address} {Cham},\ \bibinfo
  {year} {2016})\BibitemShut {NoStop}%
\bibitem [{\citenamefont {Bonitz}\ \emph {et~al.}(2018)\citenamefont {Bonitz},
  \citenamefont {Scharnke},\ and\ \citenamefont {Schl\"unzen}}]{bonitz_cpp18}%
  \BibitemOpen
  \bibfield  {author} {\bibinfo {author} {\bibfnamefont {M.}~\bibnamefont
  {Bonitz}}, \bibinfo {author} {\bibfnamefont {M.}~\bibnamefont {Scharnke}},\
  and\ \bibinfo {author} {\bibfnamefont {N.}~\bibnamefont {Schl\"unzen}},\
  }\bibfield  {title} {\bibinfo {title} {{Time‐reversal invariance of quantum
  kinetic equations II: Density operator formalism}},\ }\href
  {https://doi.org/10.1002/ctpp.201700052} {\bibfield  {journal} {\bibinfo
  {journal} {Contrib. Plasma Phys.}\ }\textbf {\bibinfo {volume} {58}},\
  \bibinfo {pages} {1036} (\bibinfo {year} {2018})}\BibitemShut {NoStop}%
\bibitem [{\citenamefont {Haug}\ and\ \citenamefont
  {Jauho}(2008)}]{haug_2008_quantum}%
  \BibitemOpen
  \bibfield  {author} {\bibinfo {author} {\bibfnamefont {H.}~\bibnamefont
  {Haug}}\ and\ \bibinfo {author} {\bibfnamefont {A.-P.}\ \bibnamefont
  {Jauho}},\ }\href@noop {} {\emph {\bibinfo {title} {Quantum Kinetics in
  Transport and Optics of Semiconductors}}}\ (\bibinfo  {publisher}
  {Springer},\ \bibinfo {address} {Berlin, Heidelberg},\ \bibinfo {year}
  {2008})\BibitemShut {NoStop}%
\bibitem [{\citenamefont {Bergmann}\ and\ \citenamefont
  {Galperin}(2021)}]{galperin_epjst_21}%
  \BibitemOpen
  \bibfield  {author} {\bibinfo {author} {\bibfnamefont {N.}~\bibnamefont
  {Bergmann}}\ and\ \bibinfo {author} {\bibfnamefont {M.}~\bibnamefont
  {Galperin}},\ }\bibfield  {title} {\bibinfo {title} {{A Green’s function
  perspective on the nonequilibrium thermodynamics of open quantum systems
  strongly coupled to baths}},\ }\href {https://doi.org/10.1140/epjs/s11734-021-00067-3} {\bibfield  {journal} {\bibinfo
  {journal} {Europ. Phys. J. Spec. Top.}\ }\textbf {\bibinfo {volume} {230}},\
  \bibinfo {pages} {859} (\bibinfo {year} {2021})}\BibitemShut {NoStop}%
\bibitem [{\citenamefont {Hermanns}\ \emph {et~al.}(2014)\citenamefont
  {Hermanns}, \citenamefont {Schl{\"u}nzen},\ and\ \citenamefont
  {Bonitz}}]{hermanns_prb14}%
  \BibitemOpen
  \bibfield  {author} {\bibinfo {author} {\bibfnamefont {S.}~\bibnamefont
  {Hermanns}}, \bibinfo {author} {\bibfnamefont {N.}~\bibnamefont
  {Schl{\"u}nzen}},\ and\ \bibinfo {author} {\bibfnamefont {M.}~\bibnamefont
  {Bonitz}},\ }\bibfield  {title} {\bibinfo {title} {Hubbard nanoclusters far
  from equilibrium},\ }\href {https://doi.org/10.1103/PhysRevB.90.125111}
  {\bibfield  {journal} {\bibinfo  {journal} {Phys. Rev. B}\ }\textbf {\bibinfo
  {volume} {90}},\ \bibinfo {pages} {125111} (\bibinfo {year}
  {2014})}\BibitemShut {NoStop}%
\bibitem [{\citenamefont {Schl\"unzen}\ \emph {et~al.}(2017)\citenamefont
  {Schl\"unzen}, \citenamefont {Joost}, \citenamefont {Heidrich-Meisner},\ and\
  \citenamefont {Bonitz}}]{schluenzen_prb17}%
  \BibitemOpen
  \bibfield  {author} {\bibinfo {author} {\bibfnamefont {N.}~\bibnamefont
  {Schl\"unzen}}, \bibinfo {author} {\bibfnamefont {J.-P.}\ \bibnamefont
  {Joost}}, \bibinfo {author} {\bibfnamefont {F.}~\bibnamefont
  {Heidrich-Meisner}},\ and\ \bibinfo {author} {\bibfnamefont {M.}~\bibnamefont
  {Bonitz}},\ }\bibfield  {title} {\bibinfo {title} {{Nonequilibrium dynamics
  in the one-dimensional Fermi-Hubbard model: Comparison of the nonequilibrium
  Green-functions approach and the density matrix renormalization group
  method}},\ }\href {https://doi.org/10.1103/PhysRevB.95.165139} {\bibfield
  {journal} {\bibinfo  {journal} {Phys. Rev. B}\ }\textbf {\bibinfo {volume}
  {95}},\ \bibinfo {pages} {165139} (\bibinfo {year} {2017})}\BibitemShut
  {NoStop}%
\bibitem [{\citenamefont {Makait}\ \emph {et~al.}(2023)\citenamefont {Makait},
  \citenamefont {Borges~Fajardo},\ and\ \citenamefont
  {Bonitz}}]{makait_cpp_23}%
  \BibitemOpen
  \bibfield  {author} {\bibinfo {author} {\bibfnamefont {C.}~\bibnamefont
  {Makait}}, \bibinfo {author} {\bibfnamefont {F.}~\bibnamefont
  {Borges~Fajardo}},\ and\ \bibinfo {author} {\bibfnamefont {M.}~\bibnamefont
  {Bonitz}},\ }\bibfield  {title} {\bibinfo {title} {Time-dependent charged
  particle stopping in quantum plasmas: testing the g1–g2 scheme for
  quasi-one-dimensional systems},\ }\href {https://doi.org/10.1002/ctpp.202300008} {\bibfield  {journal}
  {\bibinfo  {journal} {Contrib. Plasma Phys.}\ 
  {e202300008}} (\bibinfo {year} {2023})}\BibitemShut {NoStop}%
\bibitem [{\citenamefont {Bonitz}\ \emph {et~al.}(1999)\citenamefont {Bonitz},
  \citenamefont {Semkat},\ and\ \citenamefont {Haug}}]{bonitz-etal.99epjb}%
  \BibitemOpen
  \bibfield  {author} {\bibinfo {author} {\bibfnamefont {M.}~\bibnamefont
  {Bonitz}}, \bibinfo {author} {\bibfnamefont {D.}~\bibnamefont {Semkat}},\
  and\ \bibinfo {author} {\bibfnamefont {H.}~\bibnamefont {Haug}},\ }\bibfield
  {title} {\bibinfo {title} {Non-{L}orentzian spectral functions for {C}oulomb
  quantum kinetics},\ }\href {https://doi.org/10.1007/s100510050770} {\bibfield  {journal} {\bibinfo  {journal}
  {Europ. Phys. J. B}\ }\textbf {\bibinfo {volume} {9}},\ \bibinfo {pages}
  {309} (\bibinfo {year} {1999})}\BibitemShut {NoStop}%
\bibitem [{\citenamefont {Lindberg}\ and\ \citenamefont
  {Koch}(1988)}]{lindberg_prb88}%
  \BibitemOpen
  \bibfield  {author} {\bibinfo {author} {\bibfnamefont {M.}~\bibnamefont
  {Lindberg}}\ and\ \bibinfo {author} {\bibfnamefont {S.~W.}\ \bibnamefont
  {Koch}},\ }\bibfield  {title} {\bibinfo {title} {{Effective Bloch equations
  for semiconductors}},\ }\href {https://doi.org/10.1103/PhysRevB.38.3342}
  {\bibfield  {journal} {\bibinfo  {journal} {Phys. Rev. B}\ }\textbf {\bibinfo
  {volume} {38}},\ \bibinfo {pages} {3342} (\bibinfo {year}
  {1988})}\BibitemShut {NoStop}%
\bibitem [{\citenamefont {Kwong}\ \emph {et~al.}(1998)\citenamefont {Kwong},
  \citenamefont {Bonitz}, \citenamefont {Binder},\ and\ \citenamefont
  {K\"ohler}}]{kwong-etal.98pss}%
  \BibitemOpen
  \bibfield  {author} {\bibinfo {author} {\bibfnamefont {N.~H.}\ \bibnamefont
  {Kwong}}, \bibinfo {author} {\bibfnamefont {M.}~\bibnamefont {Bonitz}},
  \bibinfo {author} {\bibfnamefont {R.}~\bibnamefont {Binder}},\ and\ \bibinfo
  {author} {\bibfnamefont {H.~S.}\ \bibnamefont {K\"ohler}},\ }\bibfield
  {title} {\bibinfo {title} {Semiconductor {K}adanoff-{B}aym {Equation Results
  for Optically Excited Electron-Hole Plasmas in Quantum Wells}},\ }\href
  {https://doi.org/10.1002/(SICI)1521-3951(199803)206:1<197::AID-PSSB197>3.0.CO;2-9}
  {\bibfield  {journal} {\bibinfo  {journal} {Phys. Status Solidi B}\ }\textbf
  {\bibinfo {volume} {206}},\ \bibinfo {pages} {197} (\bibinfo {year}
  {1998})}\BibitemShut {NoStop}%
\bibitem [{\citenamefont {Špička}\ \emph {et~al.}(2005)\citenamefont
  {Špička}, \citenamefont {Velický},\ and\ \citenamefont
  {Kalvová}}]{spicka_long_2005}%
  \BibitemOpen
  \bibfield  {author} {\bibinfo {author} {\bibfnamefont {V.}~\bibnamefont
  {Špička}}, \bibinfo {author} {\bibfnamefont {B.}~\bibnamefont {Velický}},\
  and\ \bibinfo {author} {\bibfnamefont {A.}~\bibnamefont {Kalvová}},\
  }\bibfield  {title} {\bibinfo {title} {{Long and short time quantum dynamics:
  I. Between Green's functions and transport equations}},\ }\href
  {https://doi.org/10.1016/j.physe.2005.05.014} {\bibfield  {journal} {\bibinfo
   {journal} {Physica E}\ }\textbf {\bibinfo {volume} {29}},\ \bibinfo {pages}
  {154} (\bibinfo {year} {2005})}\BibitemShut {NoStop}%
\end{thebibliography}


%


\end{document}